\documentclass[reprint, amsmath,amssymb,aps]{revtex4-2}
\usepackage{graphicx}
\usepackage{dcolumn}
\usepackage{bm}
\usepackage{hyperref}
\hypersetup{
    colorlinks=true,
    linkcolor=blue,
    filecolor=blue,      
    urlcolor=blue,
    citecolor=blue
    }
\usepackage{amsmath,amssymb,amsfonts,bm}
\usepackage{ulem}
\usepackage[dvipsnames]{xcolor}

\begin{document}

\title{Event shape dependence of symmetry plane correlations using the Gaussian estimator in Pb-Pb collisions at the LHC using a multiphase transport model}
\author{Sarthak Tripathy$^{1,2}$}
\author{Suraj Prasad$^{1}$}
\author{Raghunath Sahoo$^{1}$}\email[Corresponding Author: ]{Raghunath.Sahoo@cern.ch}
\affiliation{$^1$Department of Physics, Indian Institute of Technology Indore, Simrol, Indore 453552, India}
\affiliation{$^2$Indian Institute of Science Education and Research (IISER) Pune, Pune 411008, India}

\begin{abstract}
\noindent
The study of symmetry plane correlations (SPCs) can be useful in characterizing the direction of the anisotropic emission of produced particles in the final state. The study of SPCs provides an independent method to understand the transport properties of the system formed in heavy-ion collisions. Similar to anisotropic flow coefficients, which are largely influenced by the initial spatial anisotropy, SPCs also depend upon the participant plane correlations measured using the participating nucleons of the collision overlap region. In this paper, SPCs have been studied in Pb-Pb collisions at $\sqrt{s_{\rm NN}}=5.02$ TeV using the event generator AMPT. In addition to their behaviour with the changing centrality of the collision, their event shape dependence has also been studied for the first time, using the event shape classifier transverse spherocity. The Gaussian estimator has been used to evaluate the correlations, and these have been compared to the participant plane correlations defined in an analogous way to the symmetry plane correlations, and a qualitative match has been found between them. These event-shape differentiated symmetry plane correlations can be used to deduce the presence of higher-order anisotropies in the initial energy distribution, thus giving insight into the initial geometry of the colliding system, among other applications like model development and model testing using Bayesian analyses.
\pacs{}
\end{abstract}

\date{\today}
\maketitle 

\section{Introduction}
\label{intro}

In the heavy-ion collisions at ultra-relativistic speeds, the formation of a deconfined state of partons (quarks and gluons), called the quark-gluon plasma (QGP), is anticipated~\cite{Heinz:2013th, Busza:2018rrf, ALICE:2022wpn}. QGP is transient in nature, having a lifetime of the order of $10^{-23}$ seconds~\cite{Akase:1990yd}. Thus, presently, QGP formation can only be studied through the study of various indirect signatures such as collectivity \cite{Stocker:1981zz, Ollitrault:1992bk}, jet quenching \cite{CMS:2012ulu}, charmonia suppression \cite{Brambilla:2016wgg}, strangeness enhancement~\cite{Muller:1983ed}, to name a few. QGP is theorized to behave like a perfect fluid. Studies based on anisotropic flow at the RHIC and LHC energies suggest that the ratio of specific shear viscosity to entropy density ($\eta_{s}/s$) is close to $1/4\pi$ which is the lower bound of  $\eta_{s}/s$ imposed by AdS/CFT calculations~\cite{Heinz:2005zg, Gonzalez:2020bqm, Schenke:2011bn, Gale:2012rq, Danielewicz:1984ww, Kovtun:2004de, Luzum:2008cw}. Further, for a smaller value of $\eta_{s}/s$, the hydrodynamic nature of the QGP can transform the initial state spatial anisotropy of the collision overlap region defined by the participant nucleons into the final state momentum space azimuthal anisotropy of the produced charged particles~\cite{Wong:1995jf, Ollitrault:1992bk, Voloshin:1994mz}. The final state azimuthal anisotropy of the charged particles can be quantified using the coefficients of the Fourier expansion of azimuthal distributions of the particles, as follows~\cite{Voloshin:1994mz}.

\begin{equation}
\label{flow_eqn}
f(\phi) = \frac{1}{2\pi}\left[1+2\sum_{n=1}^{\infty}v_{n}\cos[n(\phi-\psi_n)]\right]
\end{equation}
Here, $\phi$ denotes the azimuthal angle in the momentum space. The Fourier coefficients, $v_n$, are referred to as the anisotropic flow coefficients, and the parameters $\psi_n$ are called the symmetry plane angles. Together, these observables give us an idea of how the particles have been ejected. For example, the second order flow coefficient $v_2$ is called the elliptic flow~\cite{NA49:1997qey} and quantifies the extent up to which the charged particle distribution in the ($p_{x}-p_{y}$) axis can be described by an ellipse. The anisotropic flow coefficients can serve as an instrument to probe the initial state geometry of the colliding nuclei due to the collective hydrodynamic expansion of the QGP medium~\cite{Prasad:2022zbr}. Further discussions regarding the mathematical properties of the flow decomposition of the transverse momentum distribution of the final state particles can be found in Ref.~\cite{Bhalerao:2011yg}.\newline

Some trivial properties of the anisotropic flow coefficients and the symmetry plane angles corresponding to a particular harmonic can be directly inferred from Eq.~\eqref{flow_eqn} using the underlying fact that if a change of azimuthal coordinates is made from $\phi\to-\phi$, then the distribution $f(\phi)$, in Eq.~\eqref{flow_eqn} remains invariant. Due to this fact, if the Fourier sum in Eq.~\eqref{flow_eqn} were to be re-expressed in terms of complex exponential terms, then $v_n = v_{-n}$ and $\psi_{n} = \psi_{-n}$. A study of the symmetry plane angles by itself is not feasible as a function of centrality or other event classifiers. This is because, in each collision, there is a random orientation of the reaction plane~\cite{Wu:2022exl}, hence changing the symmetry planes randomly as well. This leads to an event average, $\langle \psi_{n}\rangle=0$. Thus, instead of explicitly studying $\psi_n$, it is reasonable to study the correlation between the symmetry plane angles with different event classifiers, including collision centrality. Similar to anisotropic flow coefficients, the correlations of the symmetry plane angles of different orders are sensitive to both initial state correlations and the transport properties of the system. Thus, a study of symmetry plane angles can provide an independent method to understand the initial state transport properties of the medium formed in heavy-ion collisions~\cite{ALICE:2023wdn}. In this direction, using the definition of the Gaussian estimator for the symmetry plane correlations (SPC), the correlation among the symmetry planes of different orders has been studied. The computation of symmetry plane angles and of anisotropic flow coefficients involving multi-particle correlations is given below\cite{Bhalerao:2011yg}:
\begin{align}
\label{corrln_def}
        &v_{n_1}^{a_1}v_{n_2}^{a_2}...v_{n_k}^{a_k}e^{i(a_1n_1\psi_{n_1}+a_2n_2\psi_{n_2}+...+a_kn_k\psi_{n_k})} \nonumber\\&= \langle e^{i(n_1\phi_1+n_2\phi_2+...+n_l\phi_l)}\rangle
\end{align}
In the above equation, the LHS denotes the combination of anisotropic flow coefficients ($v_n$) and symmetry plane angles ($\psi_n$). The values of the harmonics $n_i$ and indices $a_i$ are such that $a_i$ are positive and hence denote the number of times that a flow term of order $n_i$, or equivalently $-n_i$ occurs. There is kept independence in the choice of harmonics, which is necessary to define different kinds of correlations appropriately \cite{Magdy:2024ooh}, such that the effect of random orientation of the reaction plane from event to event, and the effect of trivial periodicity, i.e., Eq.~\eqref{flow_eqn} being invariant under the transformation $\psi_n\to \psi_n+\frac{2m\pi}{n}$, for integers $0\le m<n$ play no role when an event average of the symmetry plane correlations is computed. The coefficients are to be chosen so that:
\begin{equation}
\label{hrmn_n_pwr}
    \sum_{i=1}^{k} a_in_i = 0
\end{equation}
The above choice of coefficients ensures that the two caveats necessary to construct a meaningful physical observable to estimate the correlation between different symmetry planes are taken care of. The integer $l$ in the RHS of Eq.~\eqref{corrln_def} denotes the number of particles that need to be correlated in order to compute the LHS. The particle's final state azimuthal angles $(\phi_1,\phi_2,...,\phi_l)$ are such that all possible multiplets of particles, with appropriate cuts (Section~\ref{sec_cuts}) are considered and their associated harmonics $(n_1,n_2,...,n_l)$ are the same as harmonics $n_i$ which appear in the LHS of Eq.~\eqref{corrln_def}, $a_i$ times. That is, the correlation in the RHS of Eq.~\eqref{corrln_def} is computed using $a_i$ different particles associated with the harmonics $n_i$.\newline

The collective behaviour, which occurs due to hydrodynamic expansion of the QGP medium, is expected to be sensitive to the initial conditions of the system of the two colliding nuclei \cite{Csernai:2006zz}. Hence, the various flow coefficients and symmetry planes could give access to the initial configuration of the nuclei if the system has a linear response to its initial conditions. More explicitly, in such a scenario, the SPCs should exhibit similar behaviour to the initial state, participant plane correlations (PPCs), with changing centrality. Further, due to event-by-event fluctuation in the geometry of the collision overlap region, both initial eccentricity and participant plane angle can vary event-by-event. Thus, in a particular centrality class, both anisotropic flow coefficients and symmetry plane angles can have a broad range distribution. This can lead to changes in symmetry plane correlations within a centrality bin. These varying magnitudes of symmetry plane correlations can be studied using event-shape observables. Transverse spherocity~\cite{Cuautle:2014yda, ALICE:2019dfi, ALICE:2023bga} is one such event shape observable that can identify the isotropic events, dominated by soft interactions, from the jetty events dominated by hard interactions. Studies based on a multi-phase transport model show the applicability of transverse spherocity to separate events in heavy-ion collisions at the LHC energies~\cite{Prasad:2021bdq, Mallick:2021hcs, Mallick:2021wop, Mallick:2020ium, Prasad:2022zbr, Prasad:2025ezg, Prasad:2025yfj}. \newline

The necessity of studying SPCs by classifying collision events by their shapes is obvious since it could give us insight into the variation of the SPCs with the processes that lead to final particle production with a certain geometry. Also, some correlations, for example, which have been experimentally found to be zero, e.g., the correlation of $\psi_2$ and $\psi_3$ \cite{ALICE:2023wdn}, can be verified if they are always zero or happen to be so due to the averaging over events of all shapes, and actually have non-zero correlation in specific event shapes. Such a study will provide insight into the usability of different event shapes to probe into correlations amongst symmetry planes. The event shape classifier used for this purpose is transverse spherocity (hereafter referred to as spherocity or $S_0$ for brevity). Previously, the event shape dependence of SPCs involving only two planes has been studied, for example, in Ref.~\cite{Huo:2013qma, Zhao:2017yhj}. However, these studies did not use the Gaussian Estimator to evaluate the SPCs; hence, they were biased by the magnitude of the involved anisotropic flow coefficients themselves, as was pointed out in Ref. \cite{Bilandzic:2020csw}.\newline 

The paper is organized as follows. We start with a brief introduction in Section~\ref{intro}, followed by the description of the methodology of event generation, i.e., the details of a multi-phase transport model (AMPT) parameters and the phase space cuts on particle selection for SPC evaluation in Section~\ref{sec_method}.
The results are discussed in Section~\ref{sec_results}.
~Finally, the paper is summarised in Section~\ref{sec_summary}.

\section{Event Generation and Methodology}
\label{sec_method}

In this section, we briefly discuss the event generation using AMPT, transverse spherocity, event and track selection cuts, and the method to calculate the symmetry plane correlations.
\subsection{A Multi-phase Transport model (AMPT)}
\label{sec_ampt}
AMPT \cite{Lin:2004en} is a Monte Carlo (MC) based transport model that simulates heavy-ion collisions by explicitly dealing with non-equilibrium dynamics of the collision system. AMPT involves four major stages. In the string melting version of AMPT (AMPT-SM), the collision between two nuclei is initialized using HIJING \cite{Wang:1991hta,Gyulassy:1994ew}, followed by partonic scatterings calculated using Zhang's Parton Cascade (ZPC) \cite{Zhang:1997ej}, followed by hadronization using quark coalescence model \cite{Lin:2001zk}, and finally evaluating the hadronic rescatterings using a relativistic transport (ART) framework \cite{Li:1995ix,Li:1995pra}. In this study, we use AMPT to simulate Pb-Pb collisions at $\sqrt{s_{\rm NN}}=5020$ GeV. The parameters in the Lund symmetric splitting function are set to $a=0.30$ and $b=0.15$ \cite{Ma:2016fve}. The parton screening mass is set to 1.2408 $\rm fm^{-1}$~\cite{Zhang:1997ej}. Flags for shadowing, initial and final state radiations, and random orientation of the reaction plane angle were kept `on'. A total of 210,000 minimum bias Pb-Pb collision events are generated for this study using the settings mentioned above.

The $p_{\rm T}$-integrated anisotropic flow coefficients, such as $v_2$, $v_3$, $v_4$, in the string melting version of AMPT, which is used in the present study, are shown to slightly overestimate the experimental measurements~\cite{Mallick:2020ium, Prasad:2022zbr, Prasad:2025ezg}. Additionally, AMPT-SM overestimates the $p_{\rm T}$-differential $v_2$ in low to intermediate $p_{\rm T}$ regions (coalescence) and underestimates in the high-$p_{\rm T}$ region~\cite{Mallick:2022alr, Mallick:2023vgi}. AMPT has been shown to qualitatively reproduce many of the experimental features of anisotropic flow, as shown in the previous references, which makes it a good choice to study the symmetry plane correlations.

\subsection{Transverse spherocity}
\label{sec_sphero}
Transverse spherocity, which is used for event shape classification, is defined as follows \cite{Prasad:2022zbr, Prasad:2024gqq, MenonKavumpadikkalRadhakrishnan:2023cik, Prasad:2025yfj}:
\begin{align}
\label{sphero}
    S_0 = \frac{\pi^2}{4}\min_{\hat{n}}\left[\left(\frac{\Sigma_{i} |\vec{p}_{{\rm T}_i}\times\hat{n}|}{\Sigma_{i} |\vec{p}_{{\rm T}_i}|}\right)^2\right]
\end{align}
Where $\vec{p}_{{\rm T}_i}$ refers to the transverse momentum vector of the charged hadron track $i$ in an event and $\hat{n}$ is a unit vector in the plane transverse to the beam axis which is used to minimize the quantity within the square brackets, and the constant factor $\pi^2/4$, at the front is for normalization to unity. In accordance with its definition in Ref.~\cite{Prasad:2022zbr}, only those charged hadronic tracks with transverse momenta $p_{\rm T} > 0.15$ GeV/c have been used for the evaluation of spherocity. An additional constraint that has been used is that only those events have been considered for this study in which there are at least five charged hadrons satisfying the given transverse momentum cut. The extreme limits of $S_0\to0$ and $S_0\to1$ correspond to the event shape being pencil-like and isotropic, respectively. This is mainly because, in a pencil-like event, the particles emerge collimated along a particular direction, and hence, aligning $\hat{n}$ in that direction, all the terms in the numerator of Eq.~\eqref{sphero} vanish, leading to $S_0
\rightarrow 0$. Using the reverse arguments, we have that for an isotropic event, $S_0\rightarrow 1$. However, for all events, their $S_0$ values lie in between these extreme limits. Further details regarding spherocity can be found in Ref.~\cite{Prasad:2022zbr}.

\subsection{Participant and symmetry plane correlations}
\label{sec_defn}
From Eq.~\eqref{corrln_def}, we can reckon, for a single event, $v_ne^{in\psi_n}=\langle e^{in\phi}\rangle$. Now, as Pb-Pb collisions at the LHC are associated with high multiplicity, therefore, from event to event, $Re(v_n)$ and $Im(v_n)$ (where $Re(z),~Im(z)$ denote the real and imaginary part of a complex number $z$ respectively) are both expected to have a Gaussian distribution in the Argand plane~\cite{Bilandzic:2020csw}. Using this assumption, the correlation between symmetry planes, which are independent of the effect of flow coefficients, has been derived in Ref.~\cite{Bilandzic:2020csw}.
The Gaussian estimator for the symmetry plane correlations is given by~\cite{ALICE:2023wdn}:
\begin{widetext}
\begin{equation}
\label{main_eqn}
\langle \cos(a_1n_1\psi_{n_1}+a_2n_2\psi_{n_2}+...+a_kn_k\psi_{n_k})\rangle_{\rm GE}
=\sqrt{\frac{\pi}{4}}\frac{\langle{v_{n_1}^{a_1}v_{n_2}^{a_2}...v_{n_k}^{a_k}\cos(a_1n_1\psi_{n_1}+a_2n_2\psi_{n_2}+...+a_kn_k\psi_{n_k})}\rangle}{\sqrt{\left\langle{v_{n_1}^{2a_1}v_{n_2}^{2a_2}...v_{n_k}^{2a_k}}\right\rangle}}
\end{equation}
\end{widetext}
Here, the angular brackets denote event averaging over a suitable set of events in a particular event class (detailed in Table \ref{classfcn}). Further, following its definition in Ref.~\cite{Bilandzic:2020csw}, and as explained in Eq.~\eqref{hrmn_n_pwr}, the choice of harmonics and their powers is made such that $\sum_{i=1}^{k}a_in_i = 0$. An example of its use and evaluation using multi-particle correlations can be found in Ref.~\cite{ALICE:2023wdn}. \newline
For the sake of comparison to initial state geometry effects, the following definition of eccentricities and participant planes has been used, motivated from its original definition in Ref.~\cite{Bhalerao:2011yg} and subsequent constraints as established in Ref.~\cite{Bilandzic:2020csw}:
\begin{align}
    \label{in_state}
    &\epsilon_{n_1}^{a_1}\epsilon_{n_2}^{a_2}...\epsilon_{n_k}^{a_k}e^{i(a_1n_1\Phi_{n_1}+a_2n_2\Phi_{n_2}+...+a_kn_k\Phi_{n_k})} \nonumber \\
    &=\frac{\langle r_{1}^{|n_1|}r_{2}^{|n_2|}...r_{l}^{|n_l|}e^{i(n_1\tilde{\phi}_{n_1}+n_2\tilde{\phi}_{n_2}+...+n_l\tilde{\phi}_{n_l})}\rangle}{\langle r_{1}^{|n_1|}r_{2}^{|n_2|}...r_{l}^{|n_l|}\rangle}
\end{align}
In the LHS of this equation, $\epsilon_n$ are eccentricities of the participating nucleons in the collision overlap region, and $\Phi_n$, their corresponding symmetry angles, are called the participant plane angle. The RHS shows the way to compute the combination of $\epsilon_n$'s and $\Phi_n$'s. The number of particles that are involved in computing the multi-particle correlations in the RHS is $\sum_{i=1}^{k} a_i = l$. In the RHS, the angles $\tilde{\phi}_i$'s are the azimuthal angles in the coordinate space of the wounded nucleons. Correspondingly, $r_i$, which appear to the power $n_i$ in the weight factors, are the radial coordinates in the plane transverse to the beam axis. The index $i$ in $r_i$ indicates that the radial coordinate being considered is that of the wounded nucleon whose azimuthal angle is $\tilde{\phi}_i$. It is important to note that Eq.~\eqref{in_state} does not entail the correct definition of $\epsilon_1e^{i\Phi_1}$\cite{Teaney:2010vd}, as well as it fails to account for the minus signs when it comes to computing just one harmonic order of initial state anisotropies. The absence of these features, however, is not an issue for the study of participant plane correlations, as first-order harmonic has not been considered for participant correlations in this paper, and in addition to that, the choice of harmonics $n_i$ would lead to all the negative signs canceling when we account for the rotational invariance of the PPCs under the random orientation of the reaction plane from event to event. In case one wishes to study the participant plane correlations involving the first harmonic, i.e., $n=1$, an explicit $r^3$ weight factor must be used \cite{Teaney:2010vd}. Finally, in analogy to Eq.~\eqref{main_eqn}, we have the following definition of the Gaussian estimator of the PPCs:
\begin{widetext}
\begin{equation}
    \label{in_main}
    \langle \cos(a_1n_1\Phi_{n_1}+a_2n_2\Phi_{n_2}+...+a_kn_k\Phi_{n_k})\rangle_{\rm GE}
    =\sqrt{\frac{\pi}{4}}\frac{\langle{\epsilon_{n_1}^{a_1}\epsilon_{n_2}^{a_2}...\epsilon_{n_k}^{a_k}\cos(a_1n_1\Phi_{n_1}+a_2n_2\Phi_{n_2}+...+a_kn_k\Phi_{n_k})}\rangle}{\sqrt{\left\langle{{\epsilon_{n_1}^{2a_1}\epsilon_{n_2}^{2a_2}...\epsilon_{n_k}^{2a_k}}}\right\rangle}}
\end{equation}
\end{widetext}
Here, the symbols have their meanings as described previously, and the angular brackets denote event averaging in the respective pool of events, considering their centrality and $S_0$ classification.

\subsection{Event and track selection}
\label{sec_cuts}
The evaluation of SPCs and PPCs as defined in Eq.~\eqref{main_eqn} and Eq.~\eqref{in_main}, respectively involve in the RHS, averaging the quantities in the angular brackets over different events. Now, each event is classified as per its centrality, which is done using its impact parameter that goes into the input during event generation. Glauber Monte Carlo scheme \cite{Loizides:2017ack}, which is used by the HIJING algorithm during nuclear configuration initialization, is also used for centrality classification. Further to this centrality classification, each event is subdivided into an event shape class based on the value of its spherocity. There are three event shape classes in each centrality class -- low-$S_{0}$ (pencil-like) events (20\% events with the least spherocity), high-$S_{0}$ (isotropic) events (20\% events with the highest spherocity) and a pool of events with all possible spherocity values, known as $S_{0}$-int events. The event shape and centrality classifications are listed in Table~\ref{classfcn}.
\begin{table}[h]
    \centering
    \begin{tabular}{|c|c|c|c|}
        \hline
         Centrality [\%] & b (fm) & Low-$S_0$ & High-$S_0$ \\
         \hline
         0-5 & 0-3.49 & 0-0.897 & 0.958-1\\
         5-10 & 3.49-4.93 & 0-0.859 & 0.942-1\\
         10-20 & 4.93-6.98 & 0-0.810 & 0.912-1\\
         20-30 & 6.98-8.55 & 0-0.764 & 0.881-1\\
         30-40 & 8.55-9.87 & 0-0.734 & 0.869-1\\
         40-50 & 9.87-11 & 0-0.719 & 0.869-1\\
         50-60 & 11-12.1 & 0-0.718 & 0.876-1\\
         \hline
    \end{tabular}
    \caption{Centrality and spherocity classification of minimum bias Pb-Pb events with $\sqrt{s_{\rm NN}}=5.02$ TeV generated using AMPT. `b' denotes the impact parameter.}
    \label{classfcn}
\end{table}
After the events have been classified into their centrality and spherocity classes, the event averages in the numerator and denominator in the RHS of Eq.~\eqref{main_eqn} are computed one by one for these different event classes.\newline
The terms inside the angular brackets of the numerator and denominator in the RHS of Eq.~\eqref{main_eqn} are obtained for each event. These, in turn, are evaluated by averaging over the charged hadron track multiplets as per Eq.~\eqref{corrln_def}. For the purpose of averaging over these tracks, only those charged hadronic tracks in the pseudorapidity range $|\eta|<0.8$ and with transverse momenta $0.2 < p_{\rm T} < 5.0$ GeV/c are used. These cuts are in order to be consistent with experiment~\cite{ALICE:2023wdn} - lower limit owing to ALICE detector efficiency, and upper limit to remove the contribution of non-flow effects that result from jets (which is taken to be $p_{\rm T}>5.0$ GeV/c).

\section{Results and Discussion}
\label{sec_results}
Firstly, it must be noted that, in systems with high multiplicity, the anisotropic flow coefficients of harmonic order $n$ quantify the extent to which the distribution of particles in the transverse momentum space looks identical to a regular polygon with $n$ sides. The anisotropic flow coefficients typically become smaller with the increase in the order of anisotropy. So, the studies of correlations amongst the anisotropic flow coefficients provide us with insight into the development of flow in the collision system. \newline
The symmetry plane angle of a harmonic order $n$, on the other hand, denotes the orientation with respect to a reference, which characterizes the particular harmonic order of anisotropy. So, the symmetry planes need not vanish, as we go on to study the higher harmonics. Hence, a non-zero value of the SPCs would mean that the symmetry plane angles being considered are correlated among themselves, indicating in turn that the corresponding harmonic orders of flow are correlated. This is why the correlation between symmetry plane angles of different harmonic orders of anisotropy will provide independent information, than what can be obtained from the correlation between anisotropic flow coefficients of the same harmonic orders. \newline
To understand this with an example, we can consider the combination $\langle\cos[6(\psi_6-\psi_3)]\rangle_{\rm GE}$. Consider that the symmetry planes of harmonic orders $n=6$ and $n=3$ are oriented at angles $\delta_6$ and $\delta_3$, respectively. Clearly, all possible symmetry planes of order 6 are $\{\delta_6+\frac{2k\pi}{6}\}_{k=0}^5$. Thus, $6\delta_6 \in \{6\delta_6+2k\pi\}_{k=0}^{5}$. Similarly, $6\delta_3\in\{6\delta_3+4k\pi\}_{k=0}^{2}$. Hence, $\cos(6(\psi_6-\psi_3)) \in \{\cos(6(\delta_6-\delta_3)+(m+2n)2\pi)\}$ for integers $m,n$. The translation of the argument of the cosine function by an integer multiple of $2\pi$ leaves it's value invariant, and ultimately we can infer that this correlator gives the correlation between any of the orientations of the $6^{\rm th}$ and $3^{\rm rd}$ order symmetry planes in the distribution of the transverse momenta of the final state particles that emerge from of a collision. Now, further generalizing, a non-zero value of the GE of the SPCs for a linear combination of symmetry planes $\sum_{i=1}^{k}a_in_i\psi_{n_i}$, indicates that, despite event-by-event fluctuations, the orientations of the harmonics $n_i$ to their respective powers of $a_i$ are correlated. While for simple choice of harmonics, we can provide direct interpretations in terms of geometry of the event in momentum space, non-trivial behaviour of correlators with centrality has been observed for different choices of $a_i$, keeping $n_i$ the same under the constraint that $\sum_{i=1}^{k}a_in_i = 0$.

\begin{figure*}
    \includegraphics[width = 0.32\linewidth]{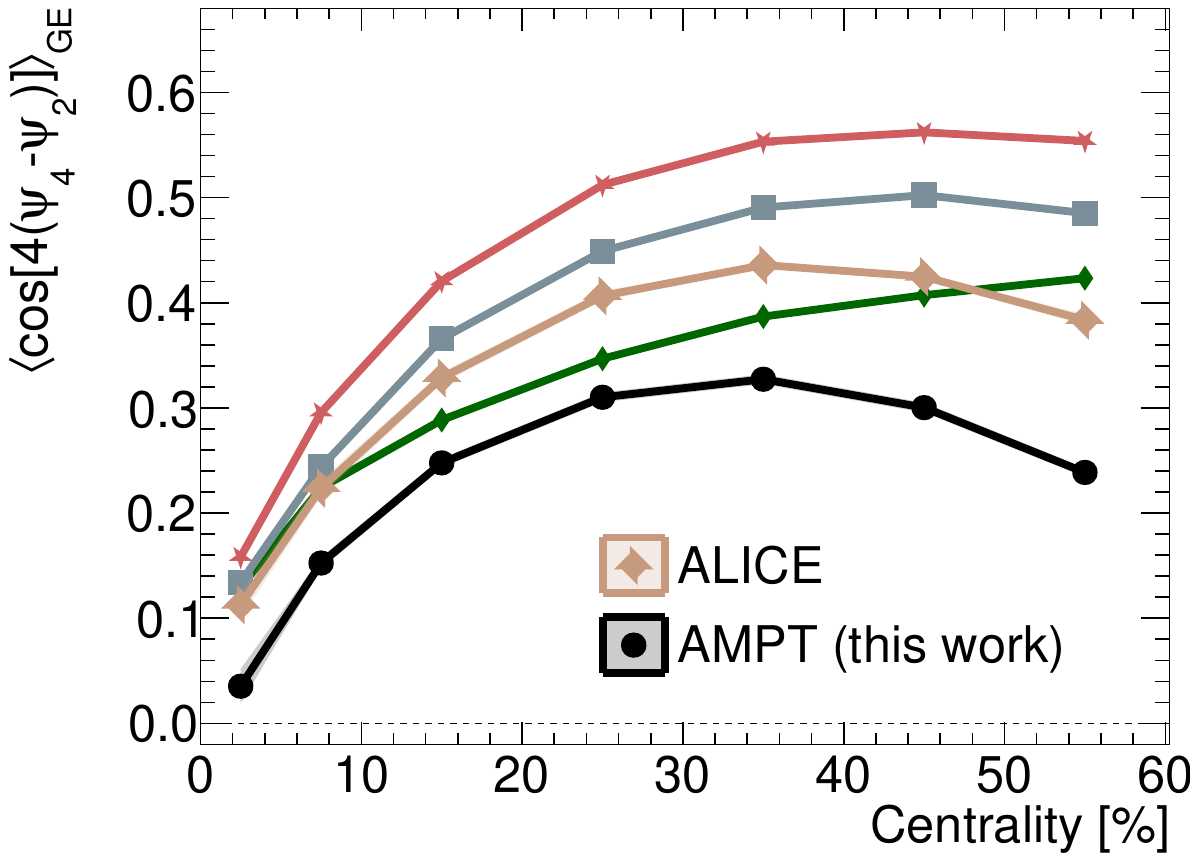}
    \includegraphics[width = 0.32\linewidth]{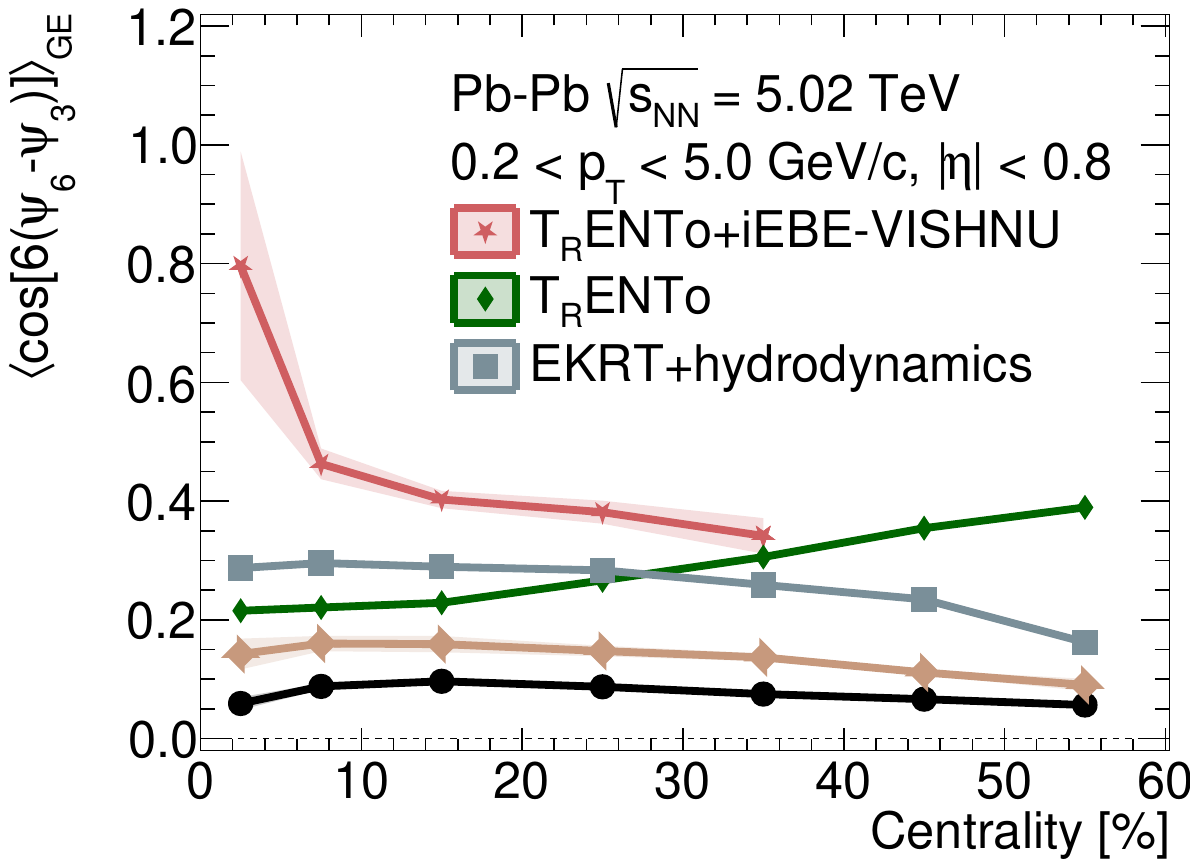}
    \includegraphics[width = 0.32\linewidth]{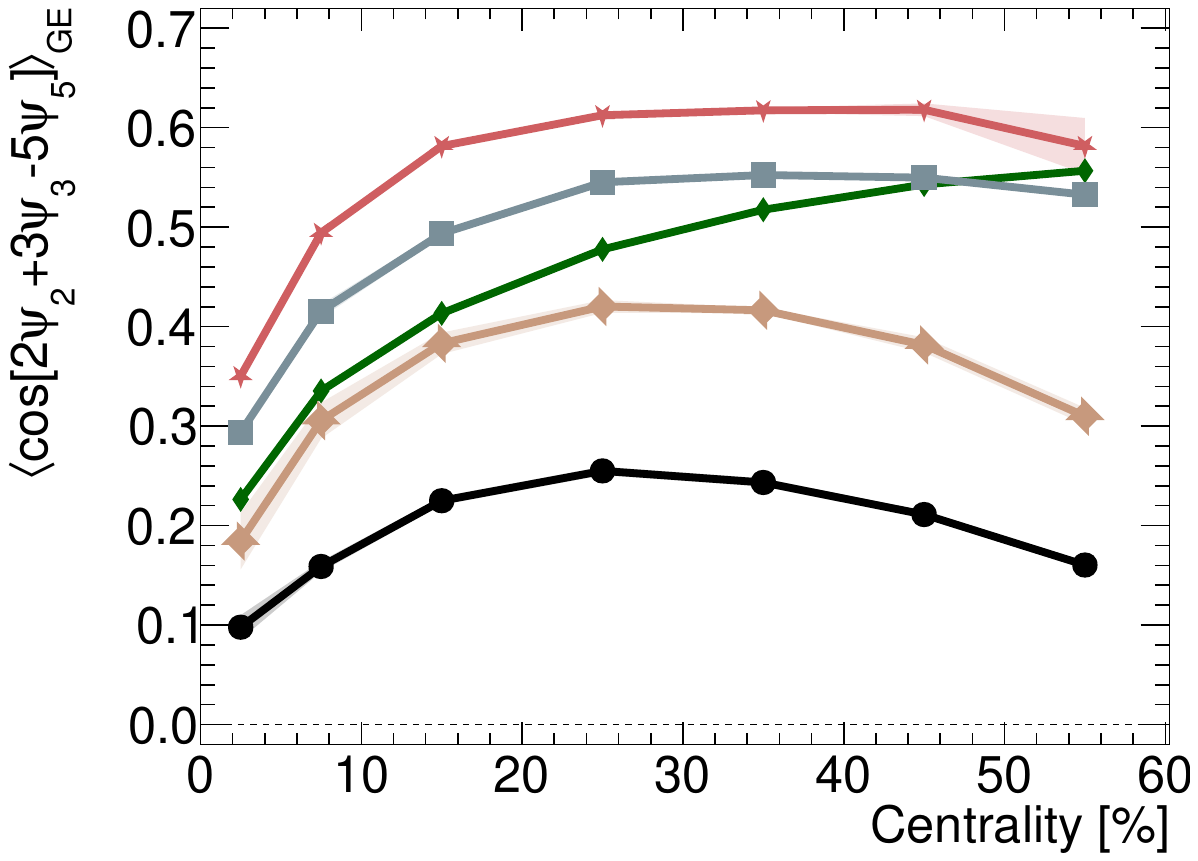}
    \includegraphics[width = 0.32\linewidth]{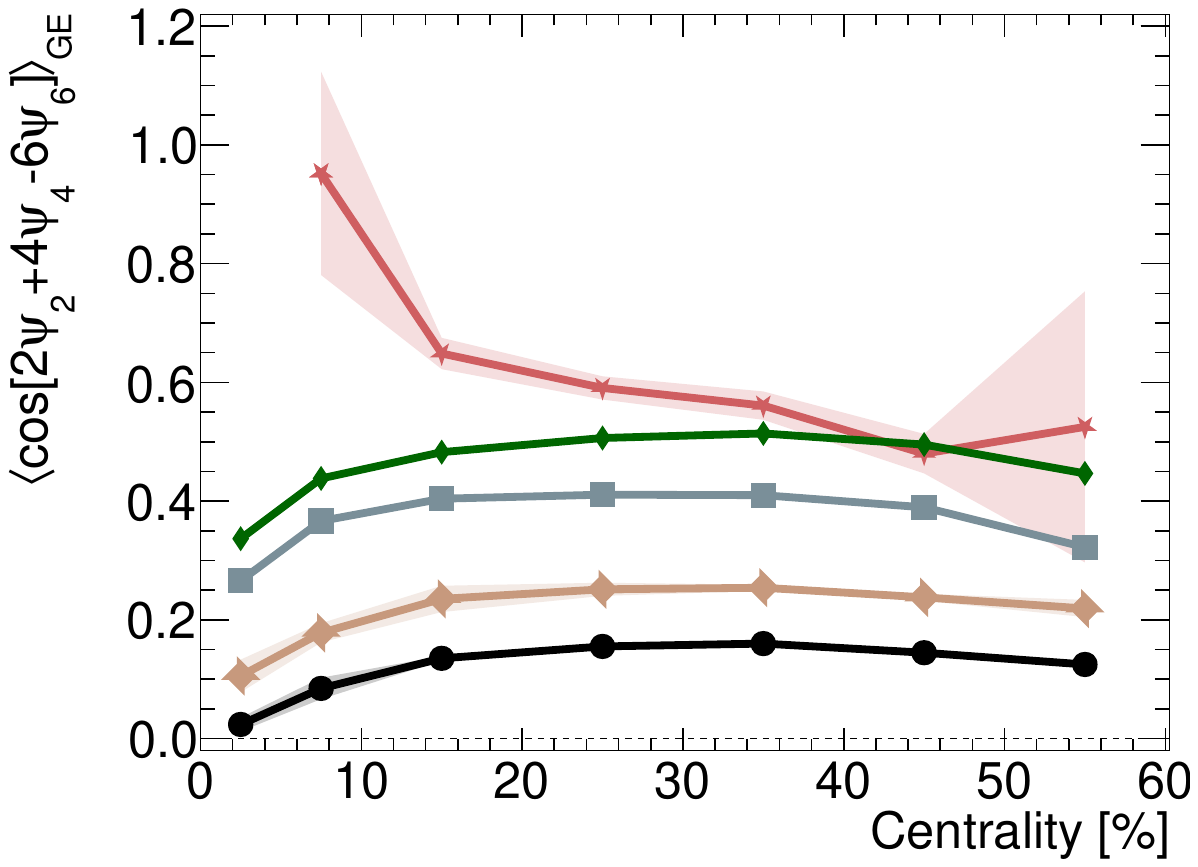}
    \includegraphics[width = 0.32\linewidth]{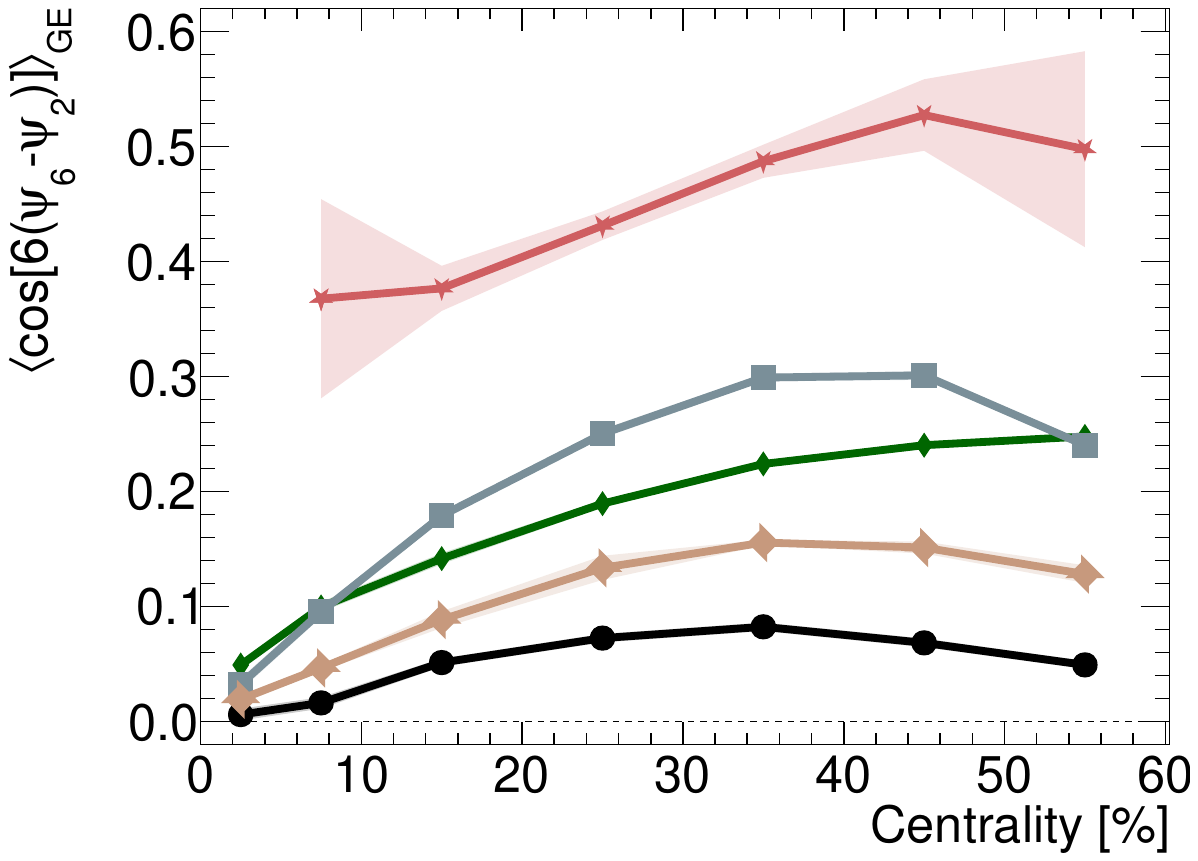}
    \includegraphics[width = 0.32\linewidth]{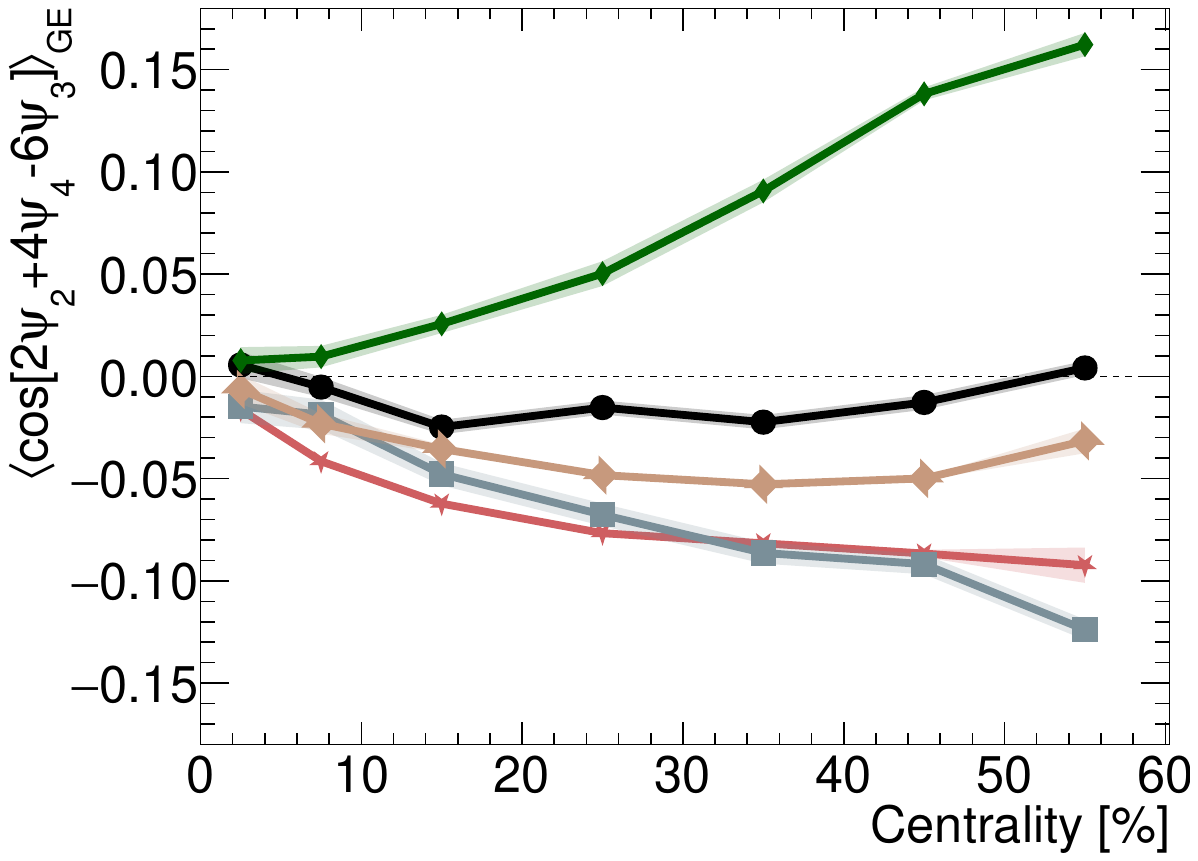}
    \includegraphics[width = 0.32\linewidth]{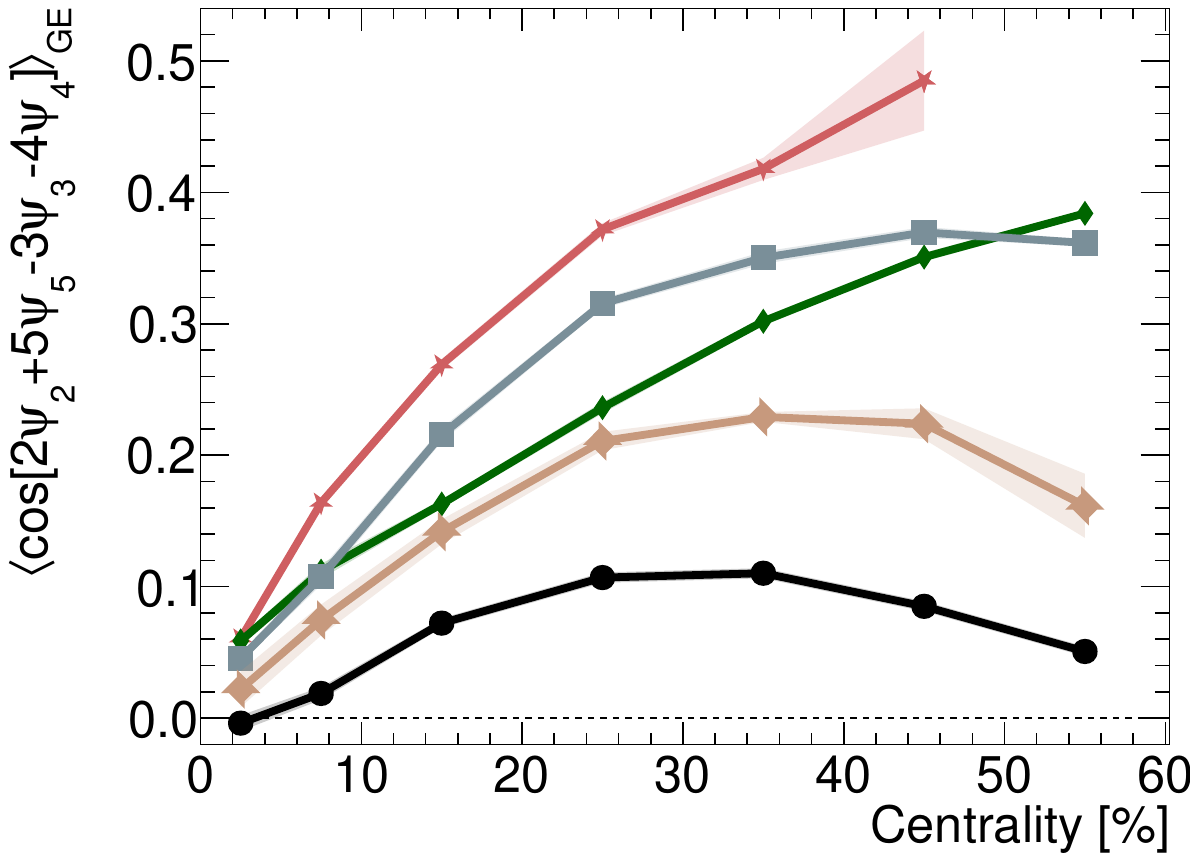}
    \includegraphics[width = 0.32\linewidth]{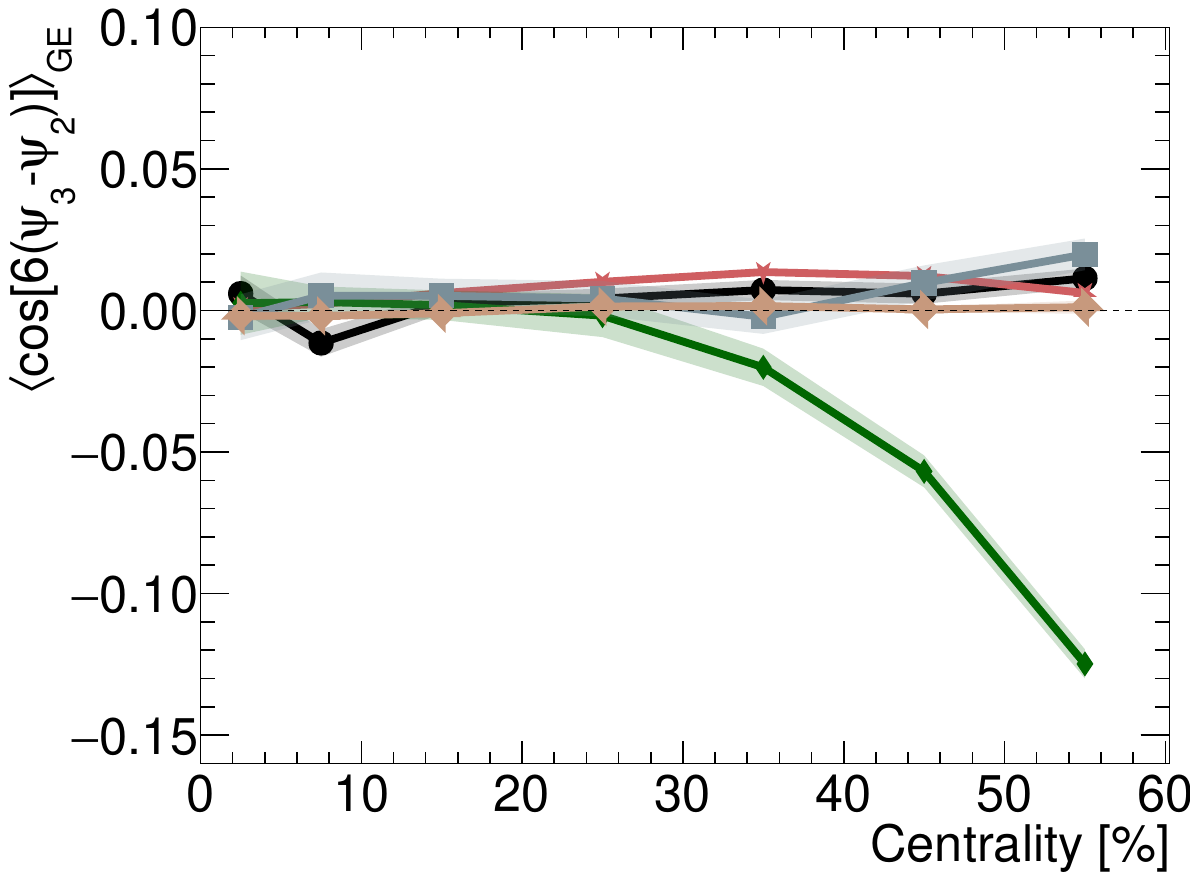}
    \includegraphics[width = 0.32\linewidth]{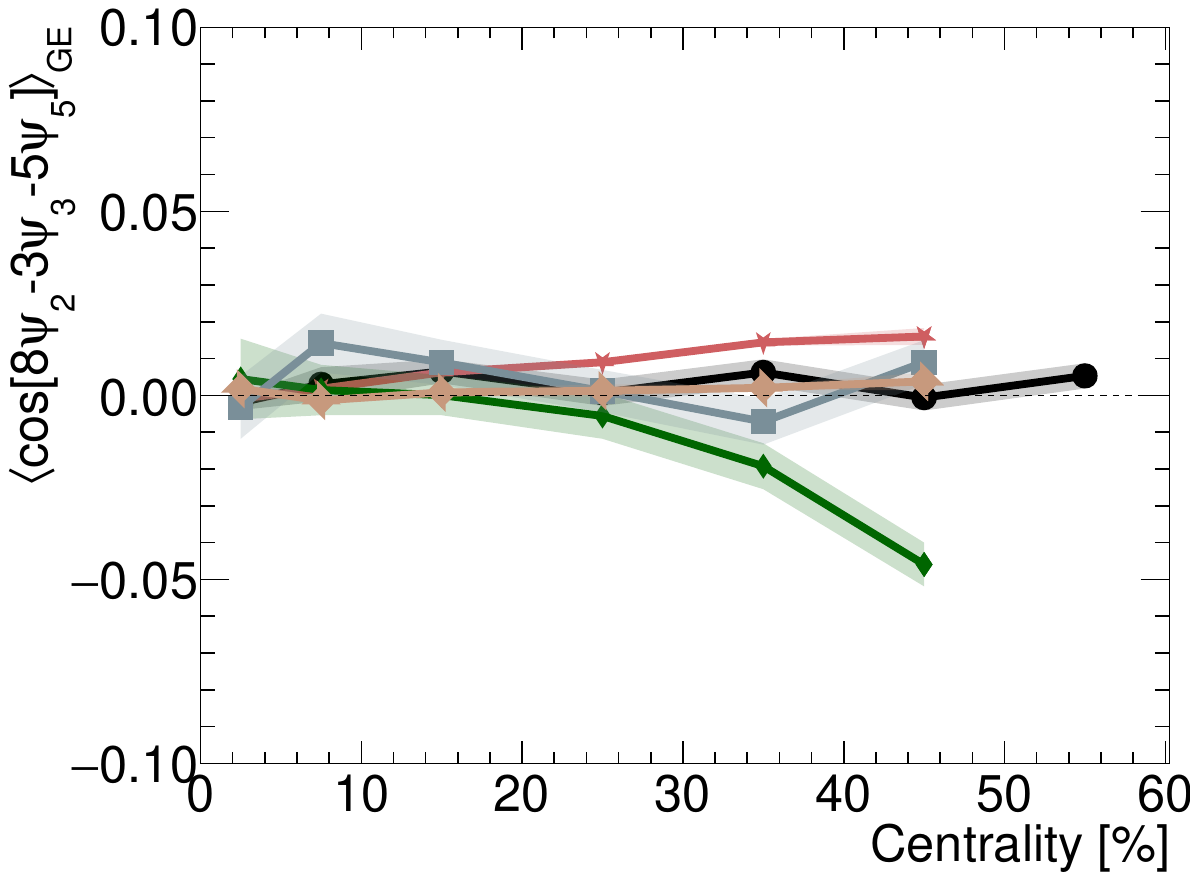}
    \caption{Comparison of the centrality dependence of SPCs calculated using various models, including AMPT from the present study, with ALICE measurements in Pb-Pb collisions at $\sqrt{s_{\rm NN}}=5.02$ TeV~\cite{ALICE:2024fus}. Corresponding measurements from T\textsubscript{R}ENTO, EKRT+hydrodynamics~\cite{Niemi:2012aj, Hirvonen:2022xfv},  T\textsubscript{R}ENTO+iEBE-VISHNU~\cite{Bass:1998ca, Bleicher:1999xi, Song:2007ux, Shen:2014vra} are taken from Ref.~\cite{ALICE:2024fus}.}
    \label{fig:compData}
\end{figure*}

\subsection{Centrality dependence of SPCs}
In this section, before explicitly discussing the SPCs, we begin with the comparison of the SPCs of the present study, calculated using AMPT, with experimental measurements from ALICE and other hydrodynamical models, shown in Fig.~\ref{fig:compData}. The measurements from T\textsubscript{R}ENTO, EKRT+hydrodynamics~\cite{Niemi:2012aj, Hirvonen:2022xfv},  T\textsubscript{R}ENTO+iEBE-VISHNU~\cite{Bass:1998ca, Bleicher:1999xi, Song:2007ux, Shen:2014vra} are taken from Ref.~\cite{ALICE:2024fus}.  Here, one finds that AMPT provides a good qualitative agreement with the experimental results while underestimating the values, while the hydrodynamical models overestimate the experimental findings. This hints that the SPCs have strong constraints on the model details.\newline

\begin{figure*}
    \includegraphics[width = 0.42\linewidth]{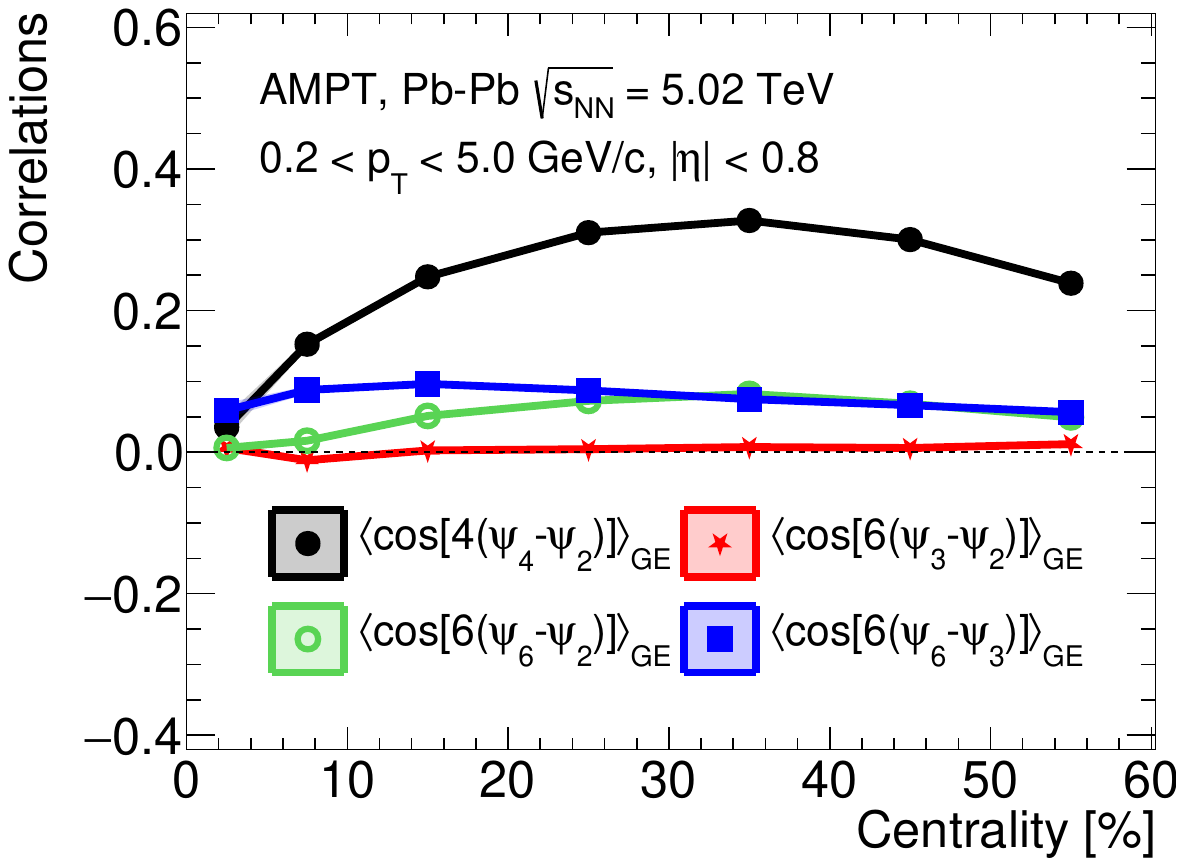}
    \includegraphics[width = 0.42\linewidth]{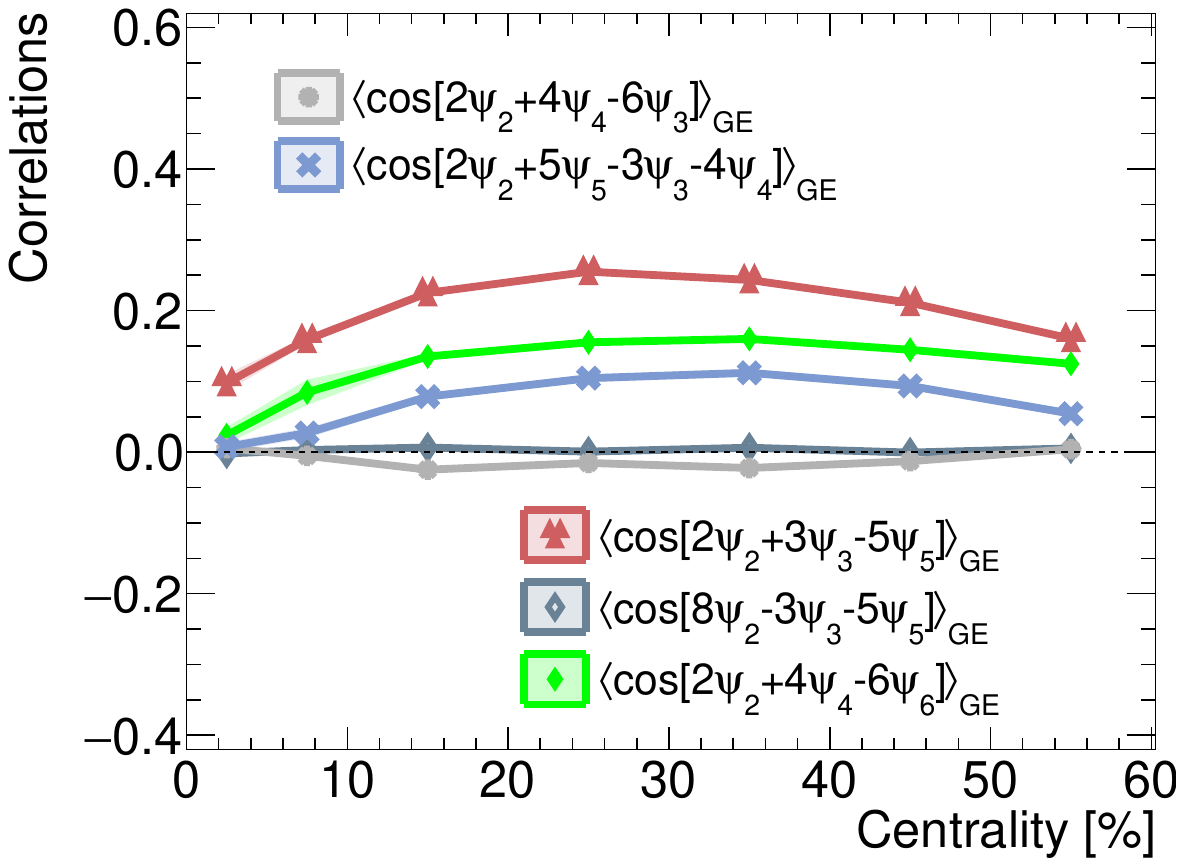}
    \caption{Centrality dependence of symmetry plane correlations of different orders in Pb-Pb collisions at $\sqrt{s_{\rm NN}}=5.02$ TeV using AMPT.}
    \label{fig:centvsSPCs}
\end{figure*}

Figure~\ref{fig:centvsSPCs} shows the symmetry plane correlations of different harmonics as a function of the centrality of collisions in Pb-Pb collisions at $\sqrt{s_{\rm NN}}=5.02$ TeV using AMPT. In the left panel, which shows the correlations for two symmetry planes, one finds that the strength of the correlation is strongest for $\langle\cos[4(\psi_4-\psi_2)]\rangle_{\rm GE}$. The correlation becomes weaker for $\langle\cos[6(\psi_6-\psi_2)]\rangle_{\rm GE}$. The correlation strength is weakest in the central collisions due to the isotropic geometry of the collision overlap region. The values of these SPCs start to rise towards the mid-central collisions, which possess contributions from initial collision geometry. Moreover, due to a smaller number of participants, the values of these SPCs start to decrease towards the peripheral collisions. 
This decreasing strength of correlation between these symmetry planes is consistent with experimental observations made in Refs.~\cite{ALICE:2023wdn, ALICE:2024fus}. \newline

Interestingly, for collision centrality (0-20)\%, one finds that $\langle\cos[4(\psi_4-\psi_2)]\rangle_{\rm GE}>\langle\cos[6(\psi_6-\psi_3)]\rangle_{\rm GE}>\langle\cos[6(\psi_6-\psi_2)]\rangle_{\rm GE}>\langle\cos[6(\psi_3-\psi_2)]\rangle_{\rm GE}$. After, (20-30)\% centrality, $\langle\cos[6(\psi_6-\psi_3)]\rangle_{\rm GE}$ and $\langle\cos[6(\psi_6-\psi_2)]\rangle_{\rm GE}$ become comparable and finally $\langle\cos[6(\psi_6-\psi_2)]\rangle_{\rm GE}$ leads the values towards (50-60)\% centrality. Although, different SPCs show different centrality dependence, $\langle\cos[4(\psi_4-\psi_2)]\rangle_{\rm GE}$ and $\langle\cos[6(\psi_6-\psi_2)]\rangle_{\rm GE}$ shows a non-linear centrality dependence. Further, the observed zero value for $\langle\cos[6(\psi_3-\psi_2)]\rangle_{\rm GE}$ indicates zero correlation between $\psi_2$ and $\psi_3$, although $v_{2}$ and $v_{3}$ show anti-correlation~\cite{ALICE:2023wdn, ALICE:2024fus, ATLAS:2015qwl, Jia:2014jca, Niemi:2015qia, Qian:2016pau, ALICE:2017kwu, ALICE:2016kpq, Niemi:2012aj, Prasad:2022zbr}. This hints towards the applicability of SPCs to independently probe the formation of a QCD medium in heavy-ion collisions.\newline

Similar to SPCs involving two symmetry planes, one finds different magnitudes for SPCs of three different symmetry planes. Here, $\langle\cos(2\psi_2+3\psi_3-5\psi_5)\rangle_{\rm GE}$ shows the strongest correlation and $\langle\cos(8\psi_2-3\psi_3-5\psi_5)\rangle_{\rm GE}$ has the weakest correlation. Interestingly, $\langle\cos(2\psi_2+3\psi_3-6\psi_6)\rangle_{\rm GE}$ has a negative sign, unlike other correlators, which are all positive or consistent with zero at the very least.\newline

Another interesting observation is that the magnitudes of the SPCs are approximately dependent on the number of particle correlations in the final state. For example, $\langle\cos[4(\psi_4-\psi_2)]\rangle_{\rm GE}$ and $\langle\cos(2\psi_2+3\psi_3-5\psi_5)\rangle_{\rm GE}$ each require only three-particle correlations and possess the largest values among the shown SPCs (Fig. \ref{fig:centvsSPCs}). In contrast, $\langle\cos[6(\psi_2-\psi_3)]\rangle_{\rm GE}$ and $\langle\cos(8\psi_2-3\psi_3-5\psi_5)\rangle_{\rm GE}$ require five and six particle correlations, respectively, and thus possess a negligibly small value~\cite{ALICE:2023wdn, ALICE:2024fus}. One possible explanation is as follows. The fluctuation of flow vectors encoded in the SPCs can be traced back to the initial state fluctuations. The initial state fluctuations can be attributed to the fluctuations due to the finite number of participants in the collision overlap region. According to the central limit theorem, the average of the random samples converges to a Gaussian distribution with an increase in the number of sampling. For a Gaussian distribution, only the second-order cumulant, namely, the width of the distribution, is non-vanishing. The skewness and kurtosis are small for a distribution close to Gaussian. Here, the order of particle correlations refers to the order of cumulants. Thus, the ordering in magnitudes of SPCs shown in Fig.~\ref{fig:centvsSPCs} indicates that the contribution of the higher order cumulant (higher order particle correlator) is smaller. On the other hand, the lowest-order cumulants have the leading values. This explains the crossing behaviour between $\langle\cos[6(\psi_6-\psi_3)]\rangle_{\rm GE}$ and $\langle\cos[6(\psi_6-\psi_2)]\rangle_{\rm GE}$, after (20-30)\% centrality class, where the number of participating nucleons is smaller.\newline
We now move to discuss the geometrical interpretation of each SPC and its corresponding PPC in the following section.

\begin{figure*}
    \includegraphics[width = 0.32\linewidth]{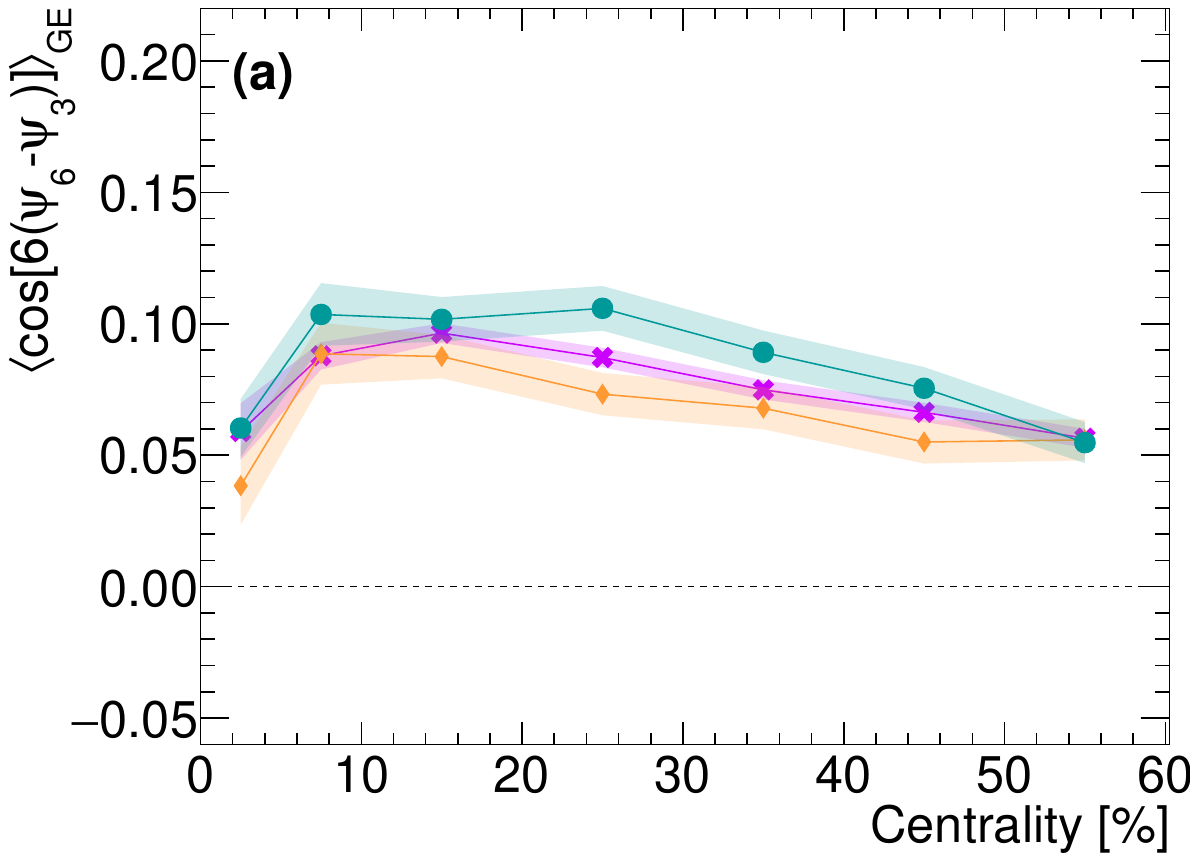}
    \includegraphics[width = 0.32\linewidth]{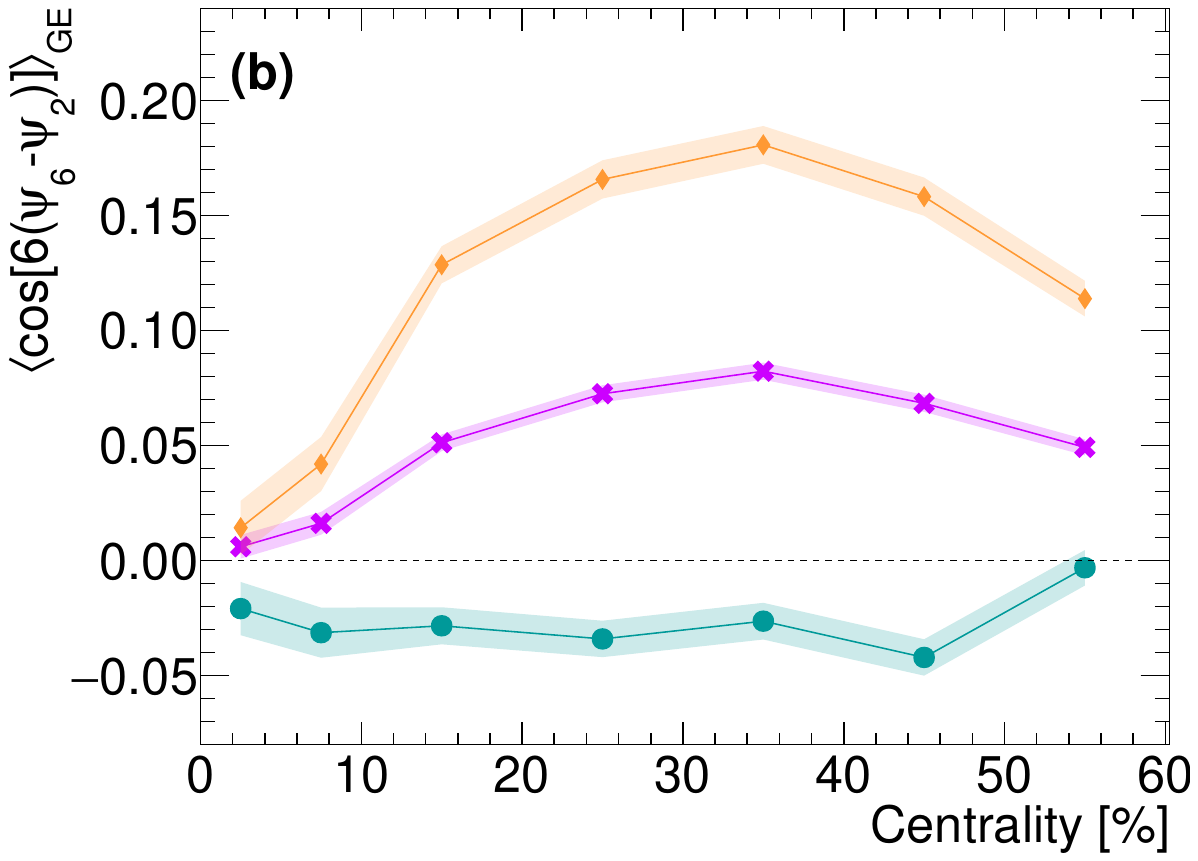}
    \includegraphics[width = 0.32\linewidth]{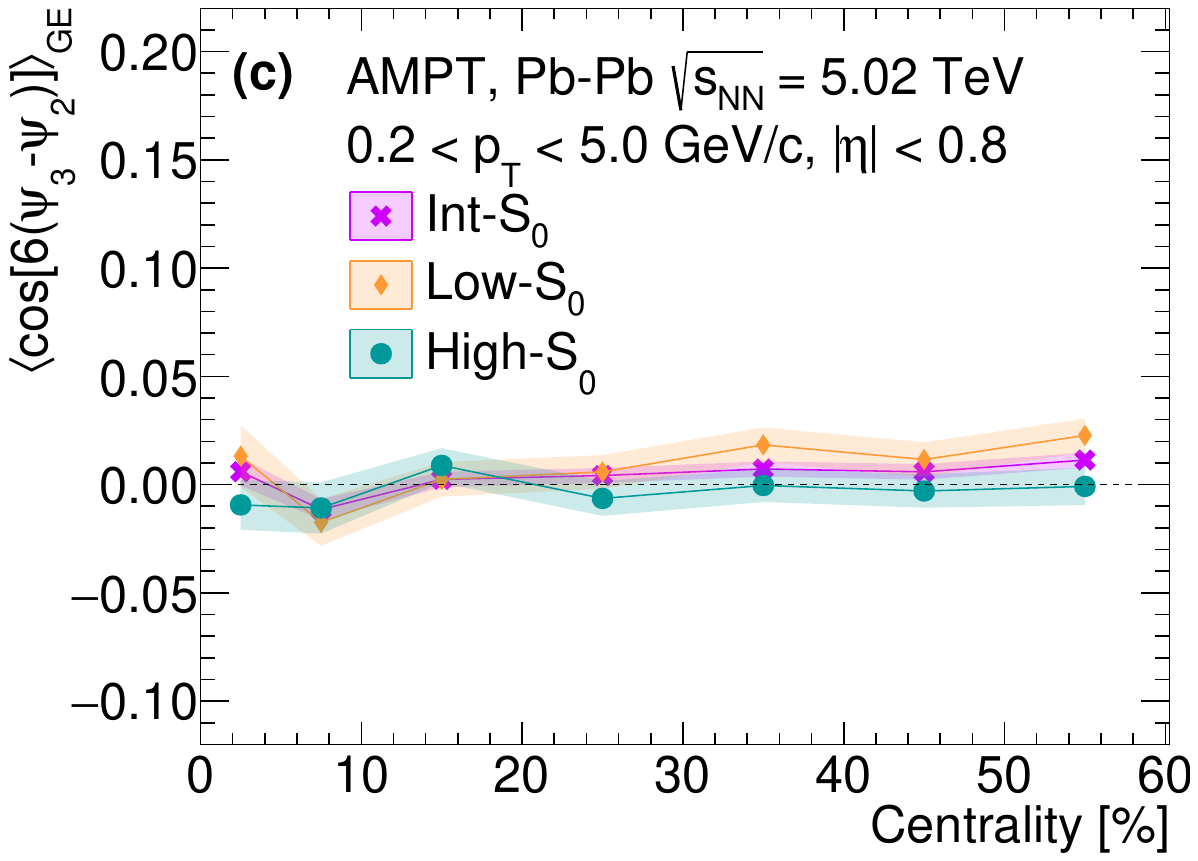}
    \includegraphics[width = 0.32\linewidth]{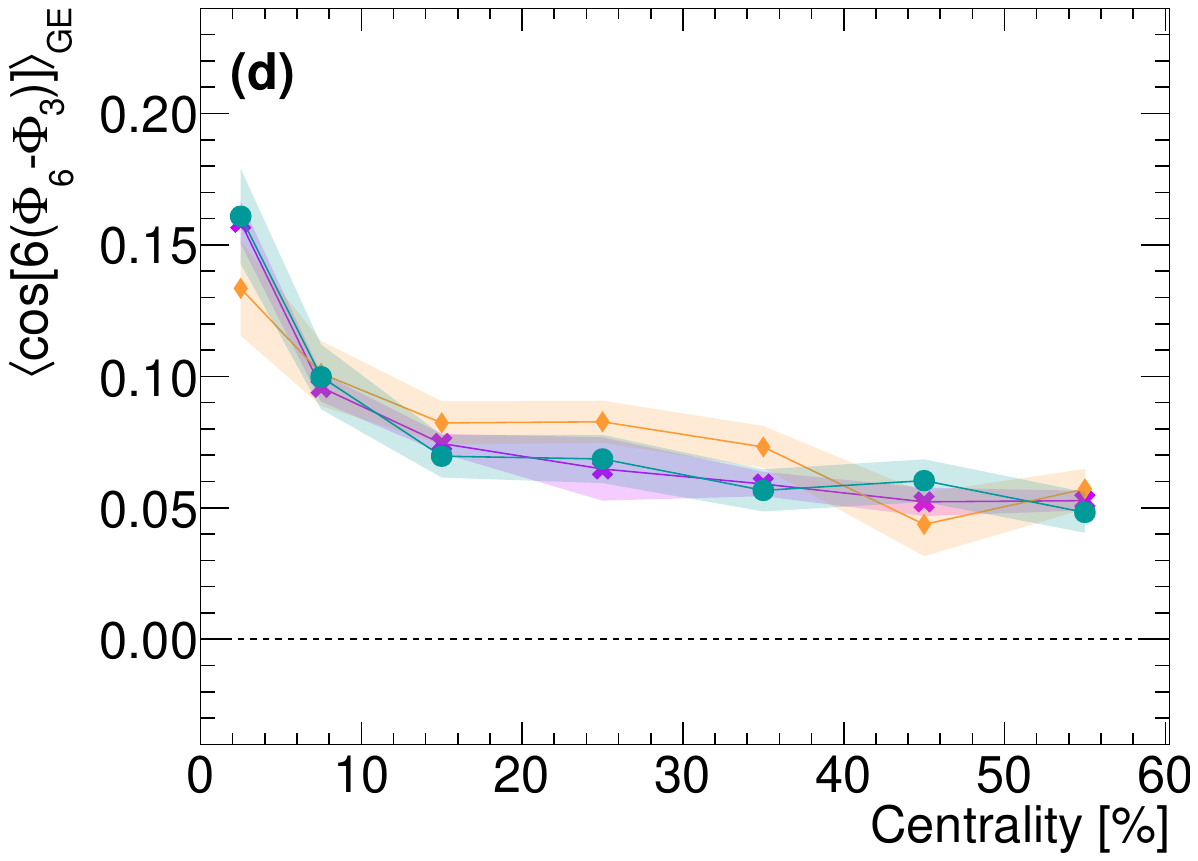}
    \includegraphics[width = 0.32\linewidth]{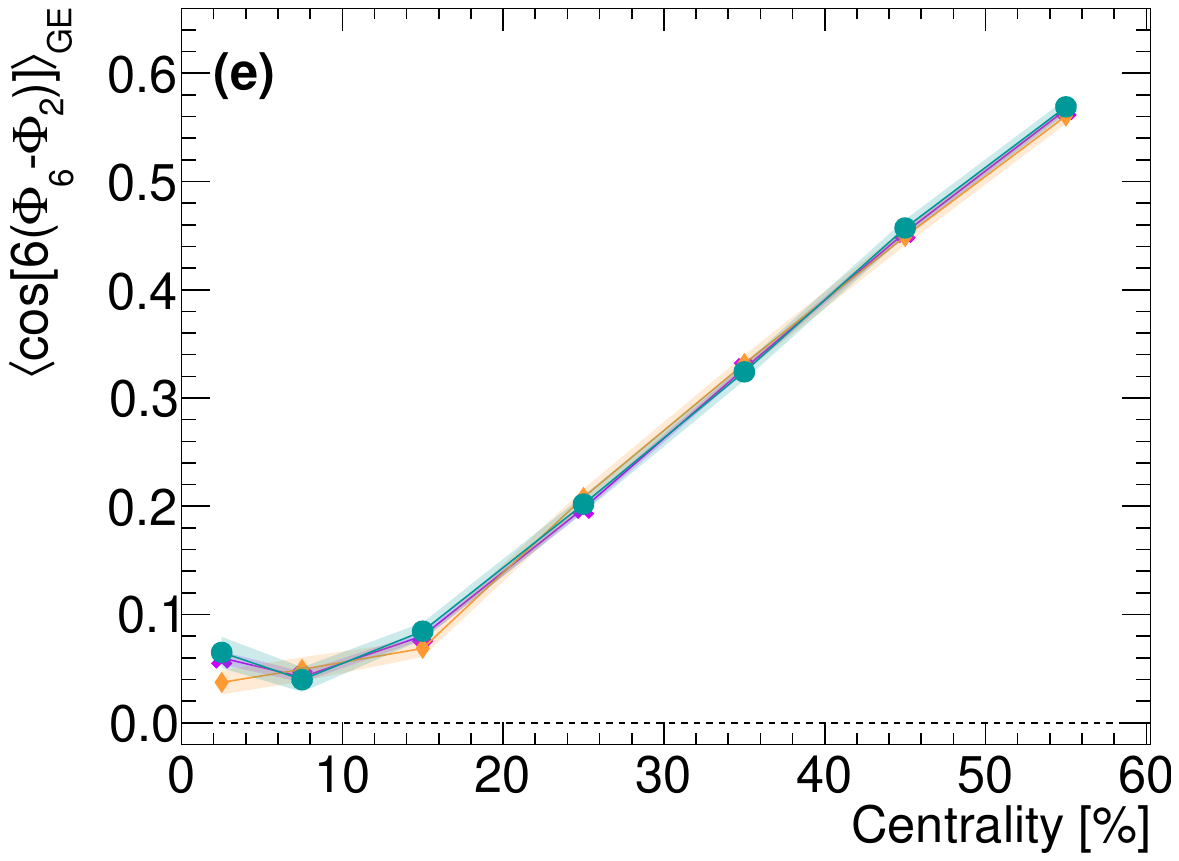}
    \includegraphics[width = 0.32\linewidth]{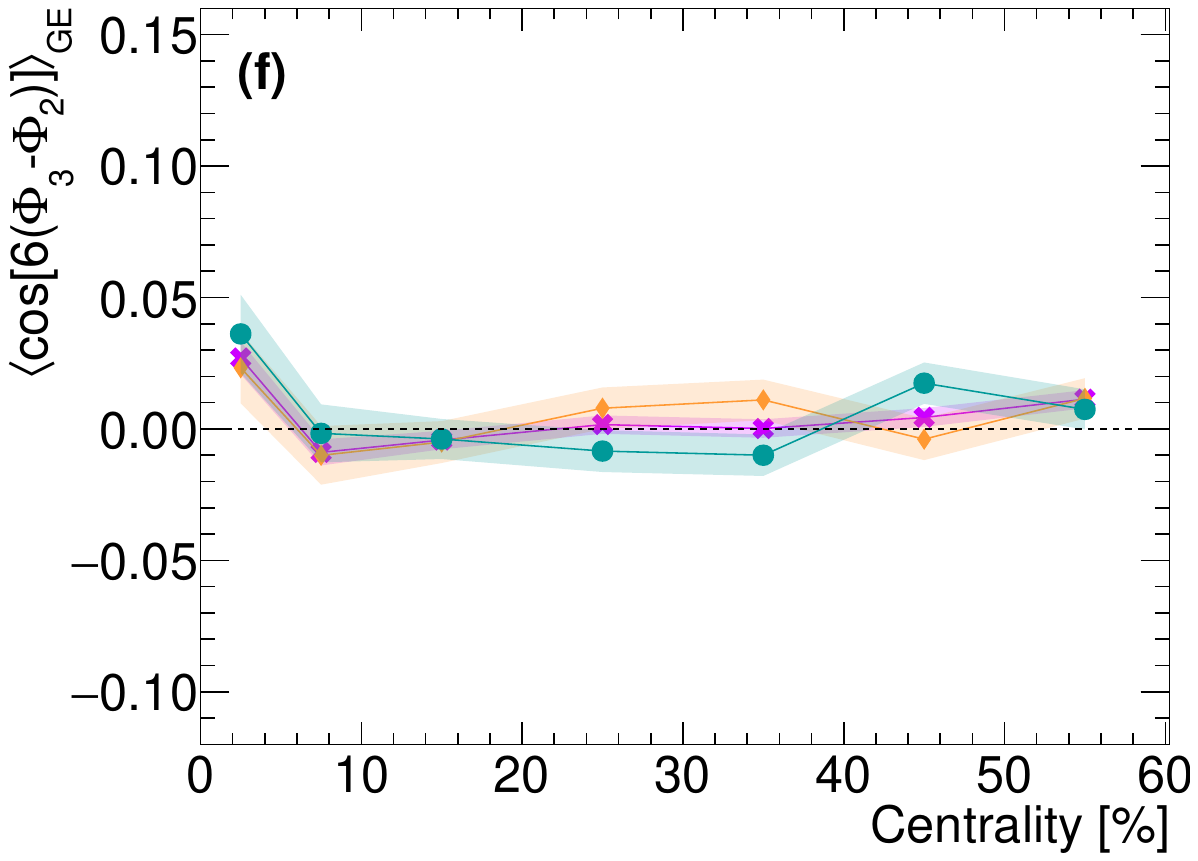}
    \caption{Centrality dependence of SPCs (shown in (a), (b), and (c)) among $n=2, 3, 6$ and corresponding PPCs (shown in (d), (e), and (f)) in Pb-Pb collisions at $\sqrt{s_{\rm NN}}=5.02$ TeV using AMPT. The calculation of SPCs considers the charged hadrons with $0.2<p_{\rm T}<5.0$ GeV/c, $|\eta|<0.8$.}
    \label{fig:SPCs236}
\end{figure*}

\subsection{Geometrical interpretation of SPCs}

\subsubsection{SPCs of harmonic orders $n=2,3,6$}
\label{236}
The symmetry planes corresponding to the harmonics $n=2,3,6$ represent elliptical, triangular, and hexagonal symmetries, respectively. The correlations amongst them have been shown in Figs.~\ref{fig:SPCs236}(a)-(c). Whenever there is a hexagonal symmetry present in the system, we can deduce trivially from the geometry that choosing any two diametrically opposite vertices of a regular hexagon (that is, those vertices that are separated by the largest distance) would also correspond to a local elliptical symmetry about them. Hence, it can be expected that there is a second-order symmetry plane whose alignment is correlated to the sixth-order symmetry plane. This gives rise to the presence of elliptic symmetry, which contributes nonlinearly in the cumulant expansion of sixth-order harmonics~\cite{Teaney:2010vd, Teaney:2012ke, Teaney:2013dta, ALICE:2024fus}. 
Similarly, by choosing any of the three vertices of a regular hexagon that alternate with each other on the hexagon, we obtain an equilateral triangle. So, a plane about which the transverse distribution has a hexagonal symmetry is also going to be correlated to a plane corresponding to triangular symmetry, i.e., the presence of non-linear triangular symmetry in the cumulant expansion of hexagonal harmonics. Thus, in addition to a strong correlation in $\psi_6$ and $\psi_2$, a strong correlation between $\psi_6$ and $\psi_3$ can also be expected. Again, along the same line of deduction, with a rare occurrence of higher $v_6$ than both $v_2$ and $v_3$ would correspond to a higher correlation between $\psi_2$ and $\psi_3$, which is not usually expected otherwise. \newline

Another interesting feature that follows from the presence of hexagonal symmetry is that there are three distinct diametrically opposite points and two distinct equilateral triangles that can be constructed using the six vertices of a regular hexagon. And more generally, given the geometrical fact that for natural numbers $m$ and $n$, we can find $m$ diameters in $2m$ sided regular polygons and $n$ equilateral triangles using the vertices of a $3n$ sided regular polygon only, it follows that in system with finite $6k$ order harmonics, for a natural number $k$, we can expect that the SPCs $\langle\cos[6k(\psi_6-\psi_2)]\rangle_{\rm GE}$ and $\langle\cos[6k(\psi_6-\psi_3)]\rangle_{\rm GE}$ would have non-vanishing magnitudes with varying centrality and spherocity. The converse will also hold true, i.e., if the SPCs $\langle\cos[6k(\psi_6-\psi_2)]\rangle_{\rm GE}$ and $\langle\cos[6k(\psi_6-\psi_3)]\rangle_{\rm GE}$ are comparable and non-vanishing, then $v_{6k}$ is non-zero in the system for some natural number $k$. This is because there exist no other regular polygons, in which a few vertices can be chosen, ensuring that the straight line connecting them is either a diameter or that they form an equilateral triangle. For the case when $k=1$, the above discussion holds true for hexagonal symmetry in the final state transverse momentum distribution.\newline

\begin{figure*}
    \includegraphics[width = 0.32\linewidth]{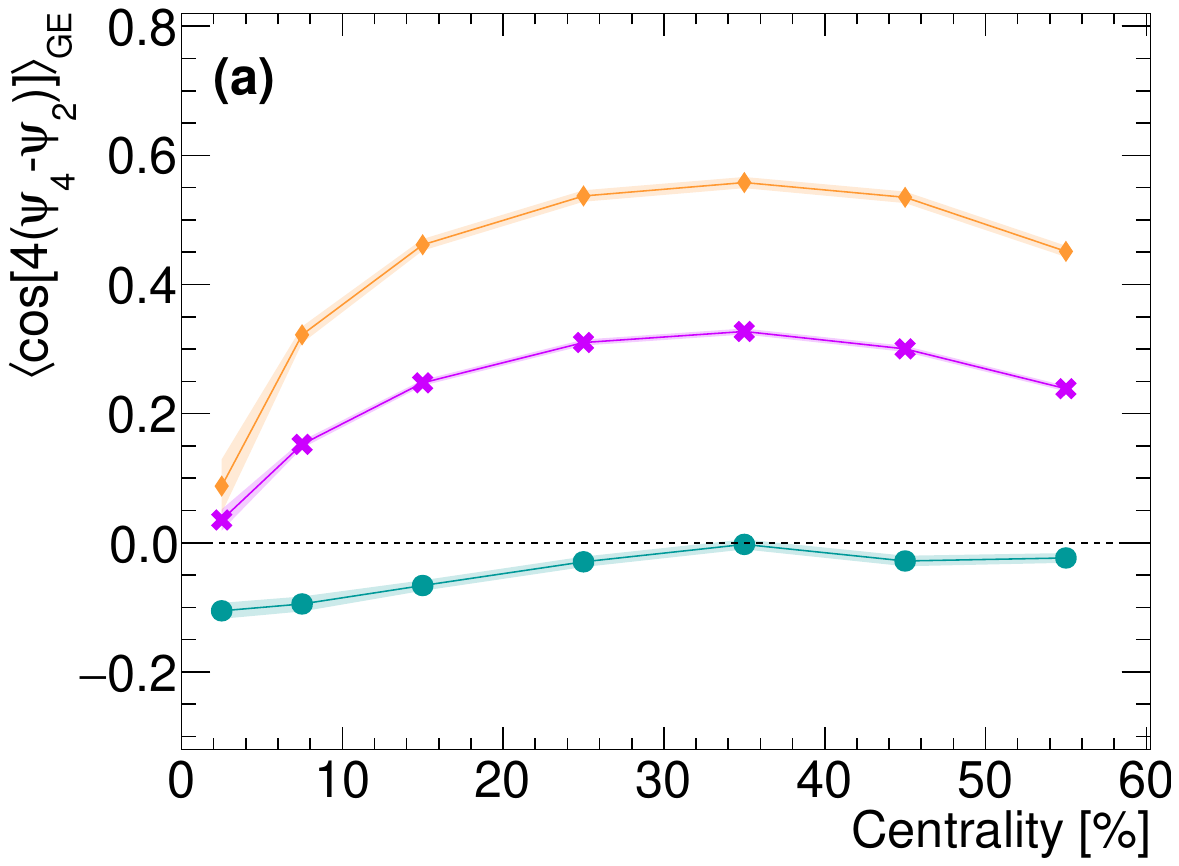}
    \includegraphics[width = 0.32\linewidth]{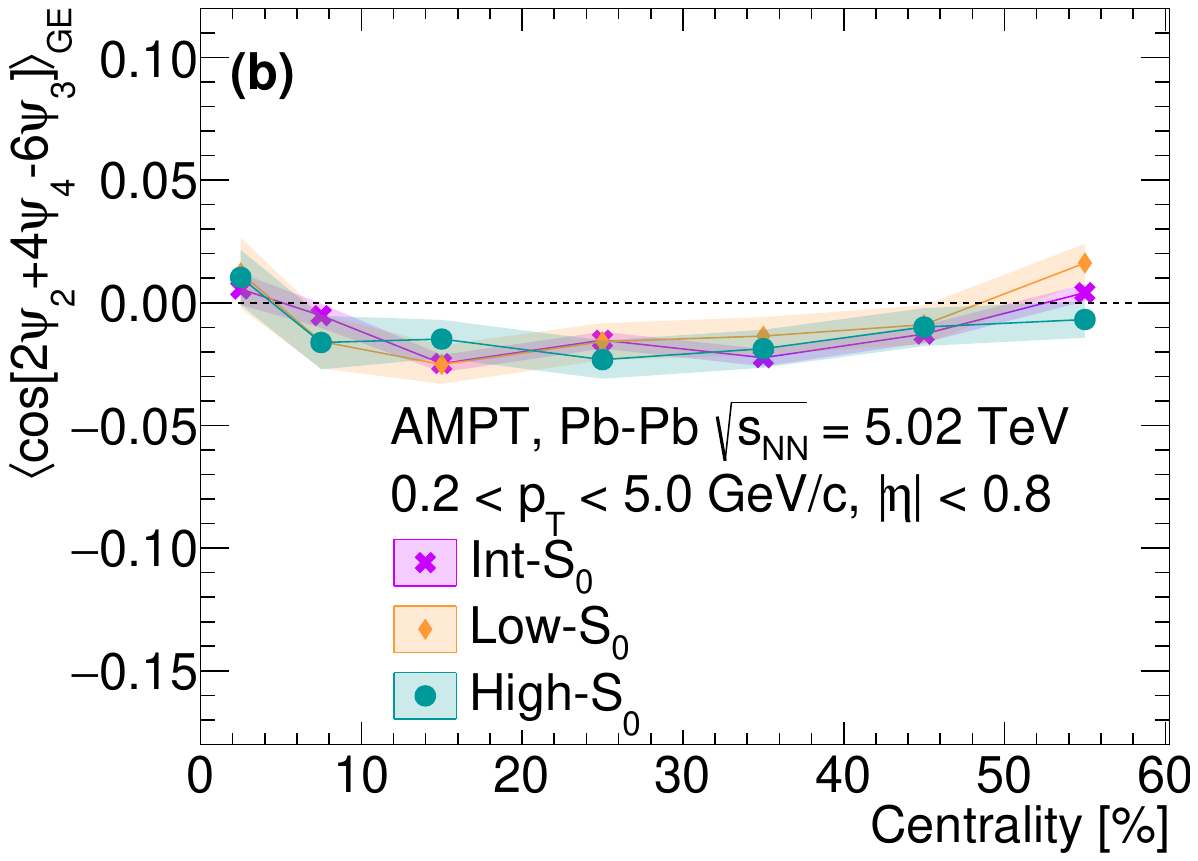}
    \includegraphics[width = 0.32\linewidth]{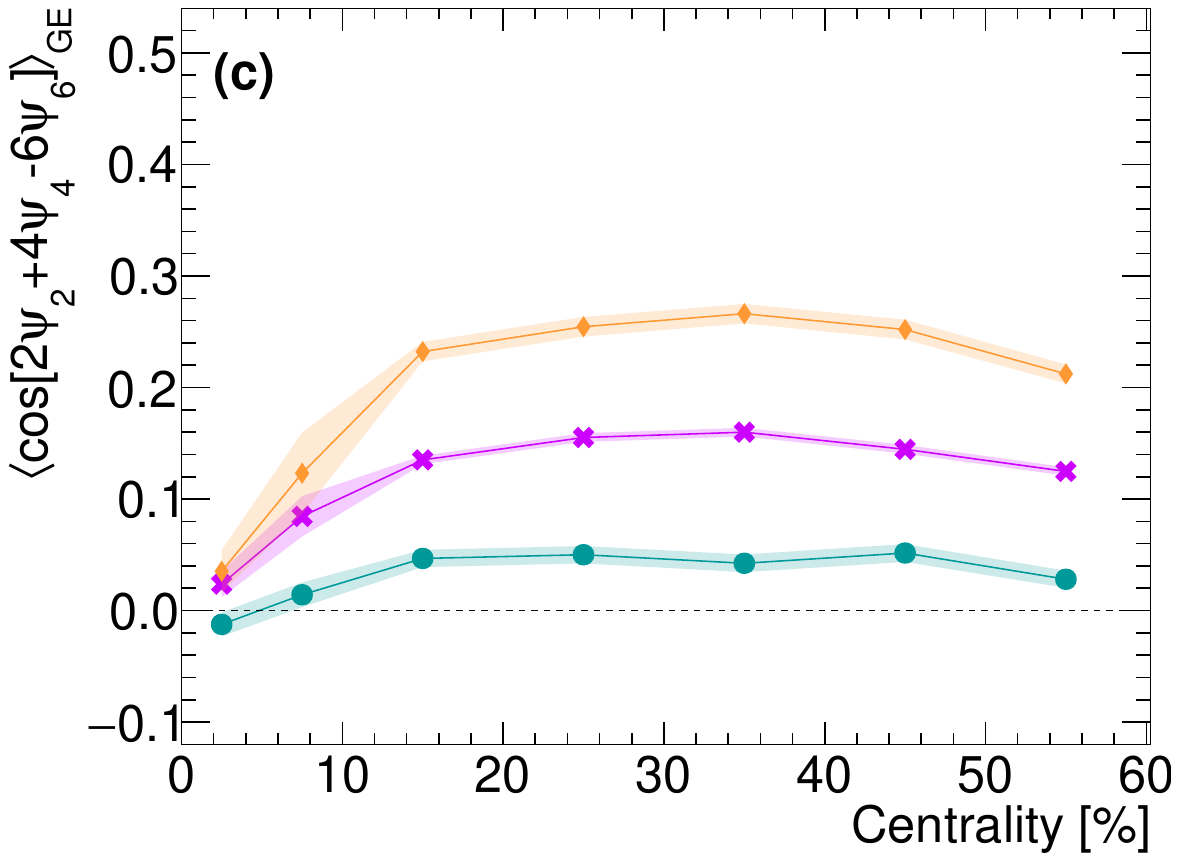}
    \includegraphics[width = 0.32\linewidth]{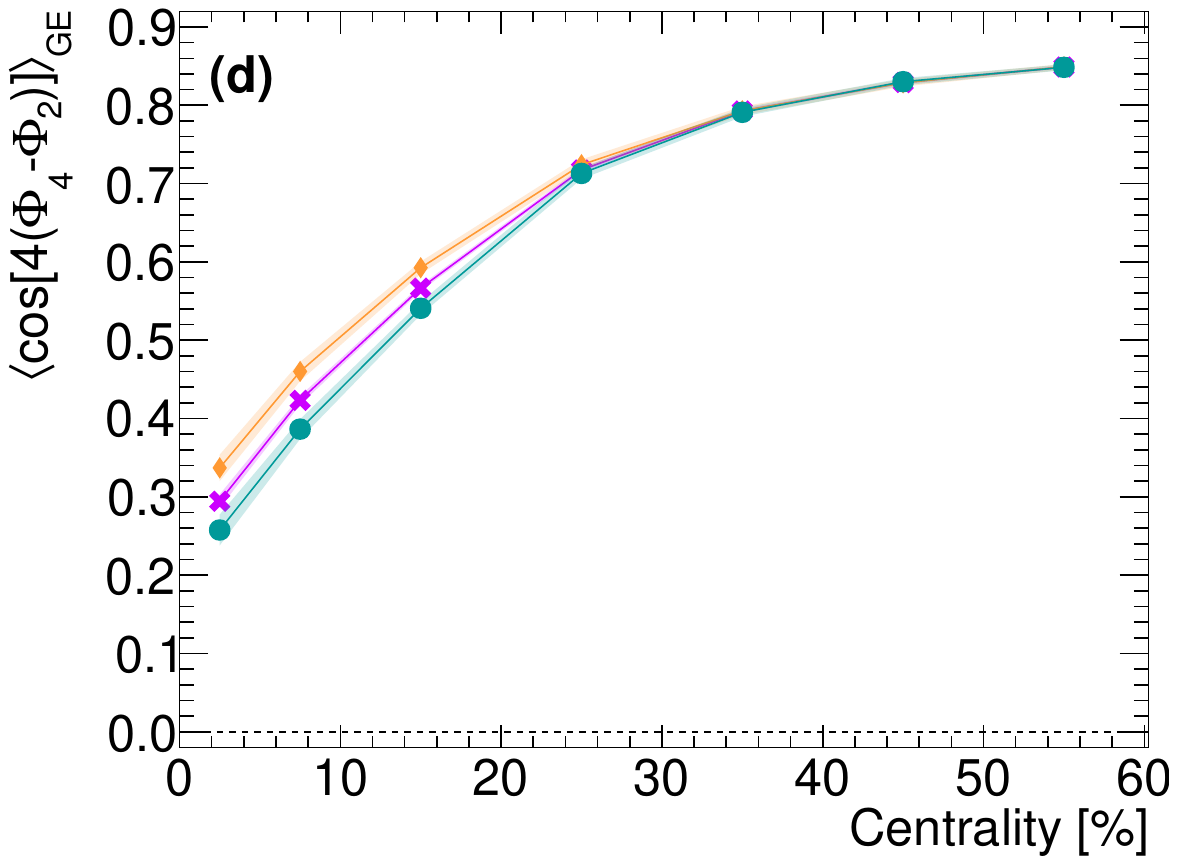}
    \includegraphics[width = 0.32\linewidth]{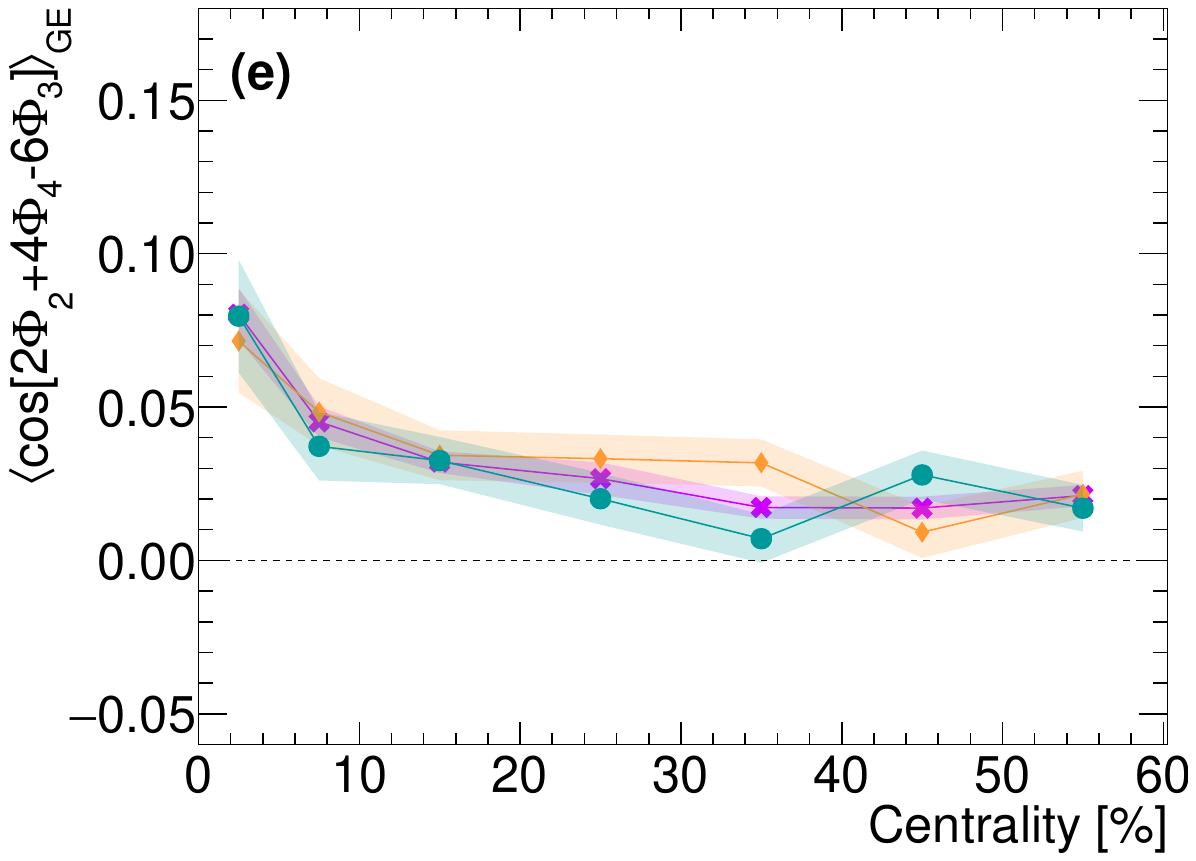}
    \includegraphics[width = 0.32\linewidth]{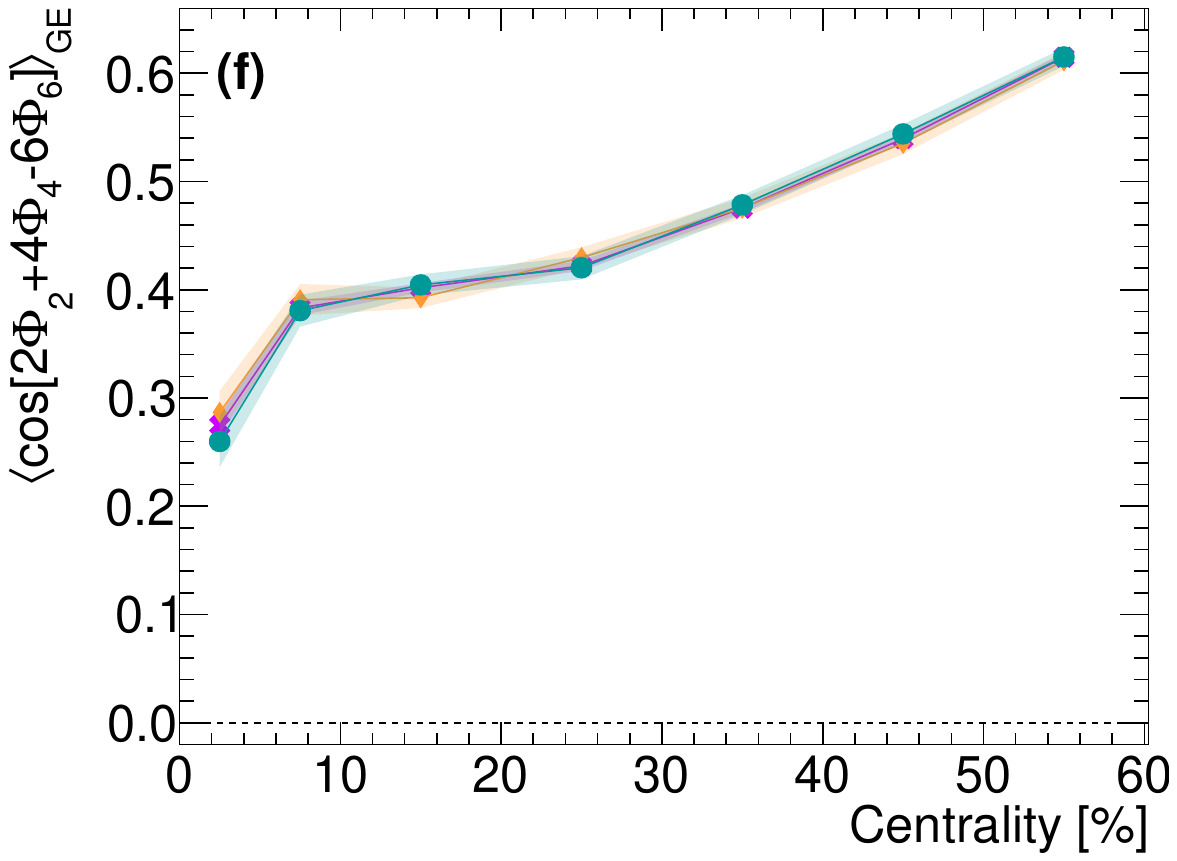}
    \caption{Centrality dependence of SPCs (shown in (a), (b), and (c)) for $n=2,4, 6$ and corresponding PPCs (shown in (d), (e), and (f)) in Pb-Pb collisions at $\sqrt{s_{\rm NN}}=5.02$ TeV using AMPT. The calculation of SPCs considers the charged hadrons with $0.2<p_{\rm T}<5.0$ GeV/c, $|\eta|<0.8$.}
    \label{fig:SPCs24}
\end{figure*}

As shown in Fig. \ref{fig:SPCs236}(a), for the most central event classes, there is an overlap in $\langle \cos[6(\psi_6-\psi_3)]\rangle$ among different event shape classes. This is indicative of the fact that this correlator takes on its values due to event-by-event fluctuations in the central Pb-Pb collisions, which is overall a reflection of corresponding PPCs shown in Fig.~\ref{fig:SPCs236} (d). More particularly, comparing Fig.~\ref{fig:SPCs24}(a) and Fig.~\ref{fig:SPCs236}(b), we can observe that although the SPC $\langle\cos[6(\psi_6-\psi_2)]\rangle_{\rm GE}$ has a qualitatively similar centrality dependence as that of $\langle\cos[4(\psi_4-\psi_2)]\rangle_{\rm GE}$ for the non-central classes, it has a different behaviour in the most central classes. From the inference previously drawn from $\langle\cos[6(\psi_6-\psi_3)]\rangle_{\rm GE}$, it can be understood that the values $\langle\cos[6(\psi_6-\psi_2)]\rangle_{\rm GE}$ for the non-central classes, is not due to an underlying anisotropy with hexagonal symmetry,  but that particle emission in Pb-Pb collisions is largely dominated by $\psi_2$, and all other harmonic terms only have a weak contribution. 
\newline

However, if second and third-order anisotropies dominate in the system, no such trivial link can be deduced. More specifically, if $v_2$ or $v_3$ dominate in comparison to $v_6$, then there are a few $6^{\rm th}$ order symmetry planes close to which the $2^{\rm nd}$ and $3^{\rm rd}$ order symmetry planes can lie in the transverse momentum space. This is precisely the case for Pb-Pb collisions at $\sqrt{s_{\rm NN}}=5.02$ TeV. Although from the above discussion, the magnitude of the anisotropic flow coefficients seems to be playing an important role in the value that the SPCs take on, it is important to note that their magnitudes have only been used to characterize the dominant symmetries of the anisotropic flow, and their contribution to SPCs has been removed by suitably choosing the harmonics and their respective powers in Eq.~\eqref{main_eqn}. \newline
\begin{figure}
    \centering
    \includegraphics[width=0.90\linewidth]{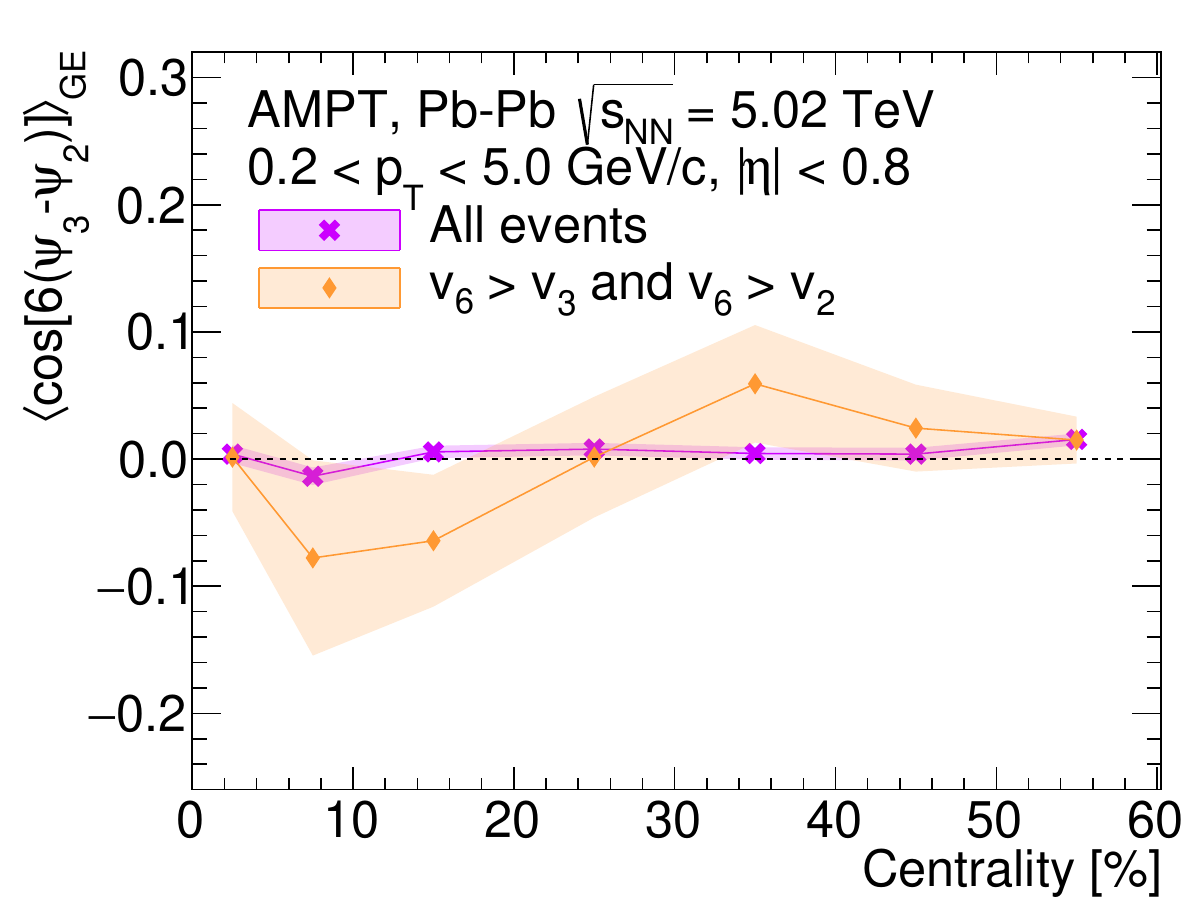}
    \caption{Comparison of SPC $\langle\cos[6(\psi_2-\psi_3)]\rangle_{\rm GE}$ between that calculated from events in which $v_6$ dominates $v_2$ and $v_3$, and all other events containing at least five charged hadrons with $p_{\rm T}>0.15$ GeV/c.}
    \label{v6_vs_v2v3}
\end{figure}

To make it clearer, a comparison has been shown in Fig.~\ref{v6_vs_v2v3} between the SPC $\langle\cos[6(\psi_3-\psi_2)]\rangle_{\rm GE}$, which has been obtained using all events in a centrality class, to that obtained using events with higher values of $v_6$. We can observe no positive shift in the values, owing to the geometrical fact that the particles are not ejected with transverse momenta distributed in a hexagonal symmetry.\newline

Gauging further into the mathematical construct of the SPCs, we find that a plethora of information about the transverse momentum distribution can be derived using SPCs of planes of lower harmonic orders. Following from the discussion at the beginning of Section~\ref{sec_results}, if hexagonal symmetry is dominant in the system, then $|\delta_6-\delta_2|\in\{0,\frac{\pi}{3},\frac{2\pi}{3}\}$. Thus, all possible values for $6(\delta_6-\delta_2)$ are of the form $2m\pi$ for some natural number $m$. Similarly, $|\delta_6-\delta_3|\in\{0,\frac{\pi}{3}\}$, and again, $6(\delta_6-\delta_3)$ will always be of the form $2m'\pi$ for some natural number $m'$. However, when considering a system where, let us say, a $12^{\rm th}$ order symmetry is dominant, there, $|\delta_6-\delta_2|\in\{0,\frac{\pi}{6},\frac{\pi}{3},\frac{\pi}{2},\frac{2\pi}{3},\frac{5\pi}{6}\}$, so that $6(\delta_6-\delta_2)\in\{0,\pi,2\pi,3\pi,4\pi,5\pi\}$. So, when these pass as arguments into the cosine function in the RHS of Eq.~\eqref{main_eqn}, some terms in the numerator would be positive while some negative, so that the average in the numerator will tend towards zero, implying in turn that the symmetry planes are decorrelated. This, however, is not the case, and in order to account for the correlation between symmetry planes of harmonic orders $n=2$ and $n=3$, in a system where a $12^{\rm th}$ order symmetry dominates, we must study the SPC $\langle\cos[12(\psi_3-\psi_2)]\rangle_{\rm GE}$.\newline

In general, if the anisotropy in the system has very high harmonic orders of symmetry planes, then that would require us to study the SPC $\lim_{k\to\infty}\langle\cos[6k(\psi_3-\psi_2)]\rangle_{\rm GE}$, which when we try to find out, starting from small natural numbers $k$, would give us vanishingly low SPC values. However, only the small $k$ values are of any physical relevance since energy occurs clustered in nuclei. It is also important to keep in mind that SPCs would invariably yield low values for an isotropic distribution of the transverse momentum vectors. In this way, a careful choice of SPCs would allow us to gauge the higher-order anisotropies of the system. It is also necessary to emphasize, however, that the errors that would be associated with finding out the dominance of higher-order symmetries in the anisotropy of the distributions are not much lesser than what we would expect by computing the anisotropic flow coefficients themselves. This is because the SPC would require a larger number of particles to be correlated as well. The usefulness of using the SPCs, however, is that we can quickly reach a conclusion about the symmetry of the dominant anisotropic flow in the system, rather than by computing each anisotropic flow coefficient and comparing them individually.


\begin{figure}
    \centering
    \includegraphics[width=0.90\linewidth]{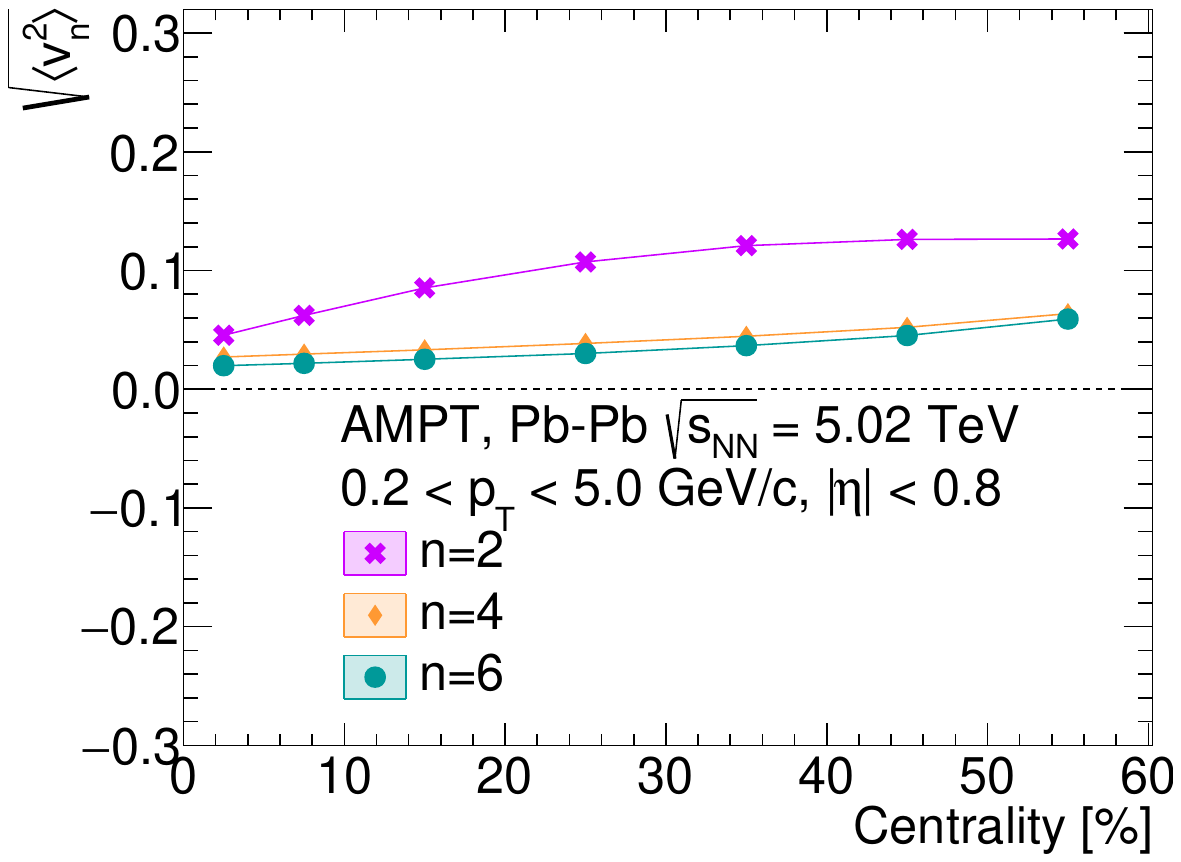}
    \caption{Comparison between the RMS values of anisotropic flow coefficients of harmonic orders $n=2,4$ and 6 in different centrality classes in Pb-Pb collisions at $\sqrt{s_{\rm NN}}=5.02$ TeV with events generated using AMPT.}
    \label{v2v4v6}
\end{figure}

\subsubsection{SPCs of harmonic orders $n=2,4,6$}
The fourth-order anisotropy would make the transverse momentum space geometry of the event appear like a square. Also, for a square, we have two distinct diagonals. These diagonals correspond to a second-order symmetry, that is, one corresponding to elliptic flow. Hence, a dominant presence of $v_4$ would imply that the symmetry planes $\psi_4$ and $\psi_2$ are strongly correlated. Hence, with decreasing centrality, the correlation between $\psi_4$ and $\psi_2$ is expected to increase, as has also been seen in Figs.~\ref{fig:SPCs24}(a) and (c). For the corresponding comparison, a plot of $v_2$, $v_4$ and $v_6$ is shown in Fig.~\ref{v2v4v6}. Furthermore, when $6^{\rm th}$ order anisotropy is dominant, it has already been described in Section~\ref{236} that the correlation between $\psi_6$ and $\psi_2$ can be expected to be strong. The presence of strong hexagonal symmetry also brings with it the presence of an approximate quadrangular symmetry and, in turn, an elliptical symmetry of the anisotropy of the transverse momentum distribution of the final charged hadrons. Thus, in the presence of high $v_4$ and $v_6$, it can be expected that the symmetry planes of orders $n=2,4,6$ are well correlated, as can be seen from the comparison of Fig.~\ref{v2v4v6} and Figs.~\ref{fig:SPCs24}(a) and (c).

\begin{figure*}
    \includegraphics[width = 0.32\linewidth]{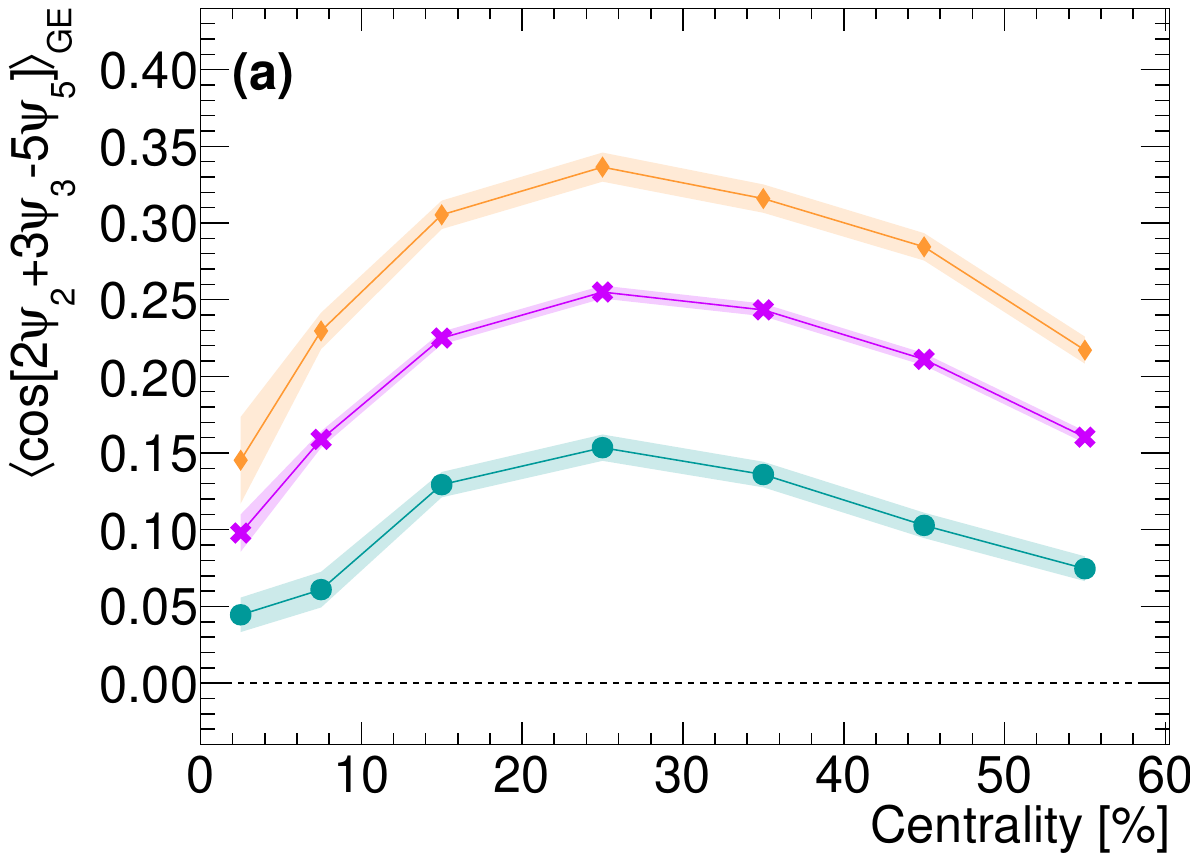}
    \includegraphics[width = 0.32\linewidth]{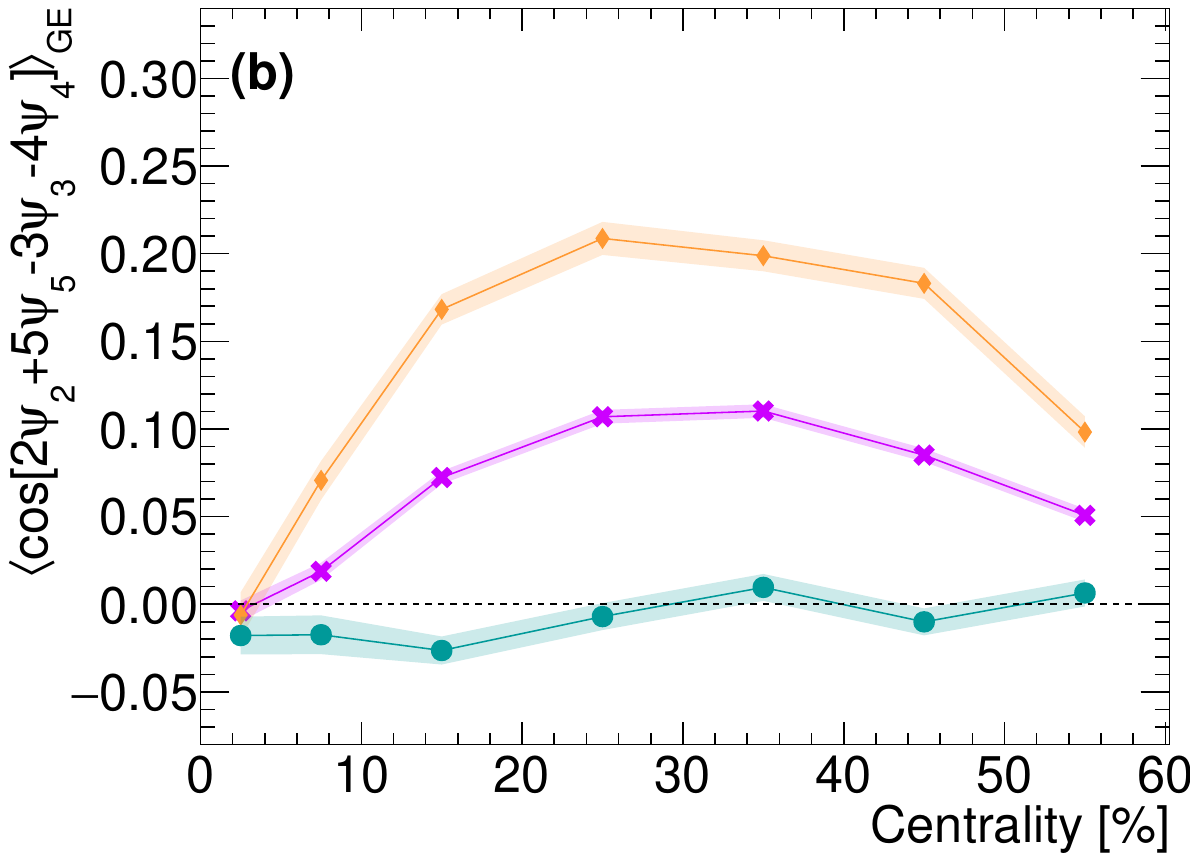}
    \includegraphics[width = 0.32\linewidth]{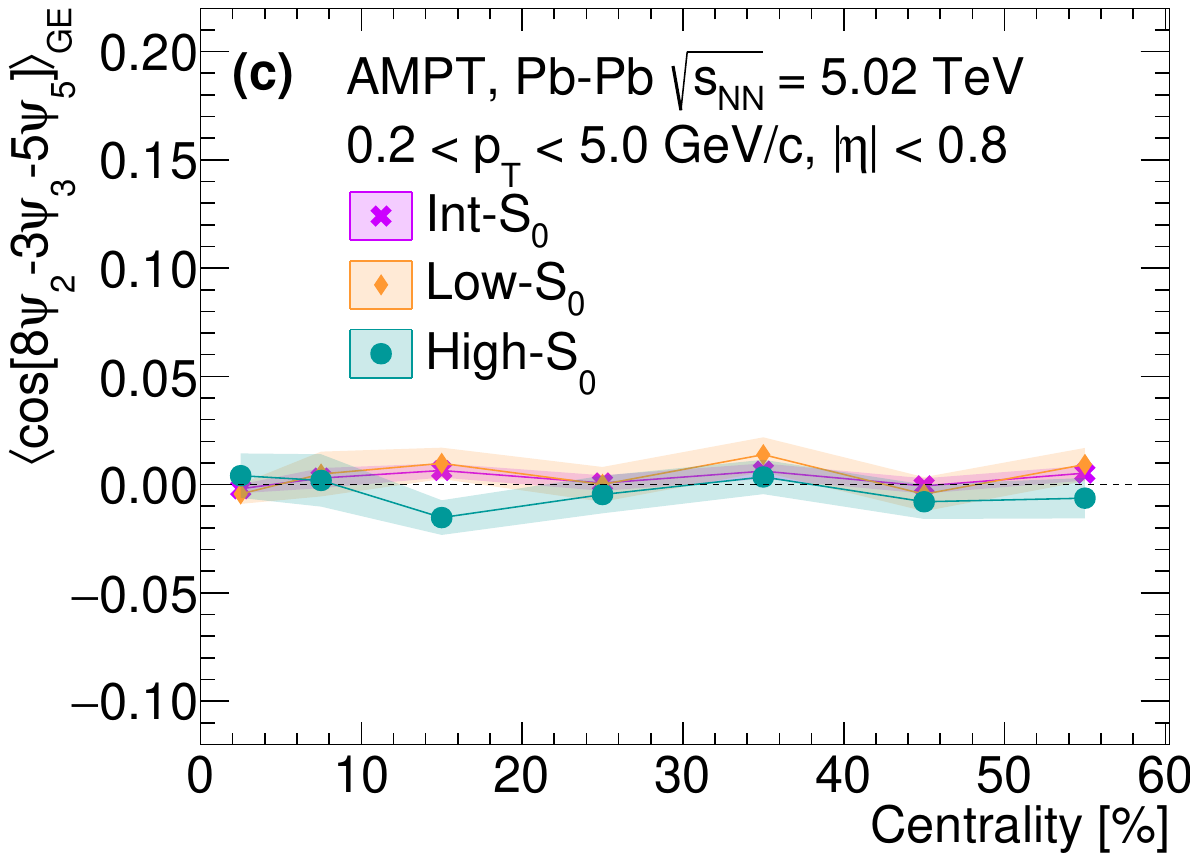}
    \includegraphics[width = 0.32\linewidth]{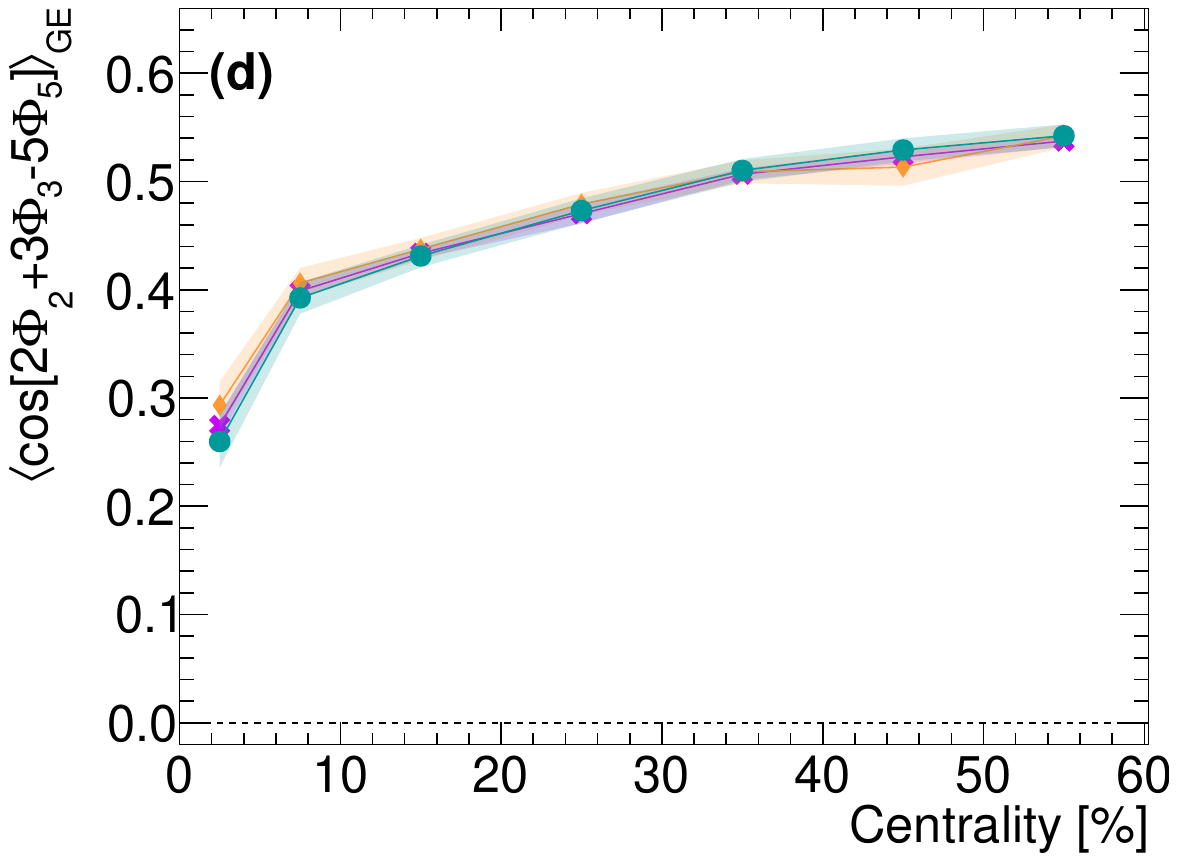}
    \includegraphics[width = 0.32\linewidth]{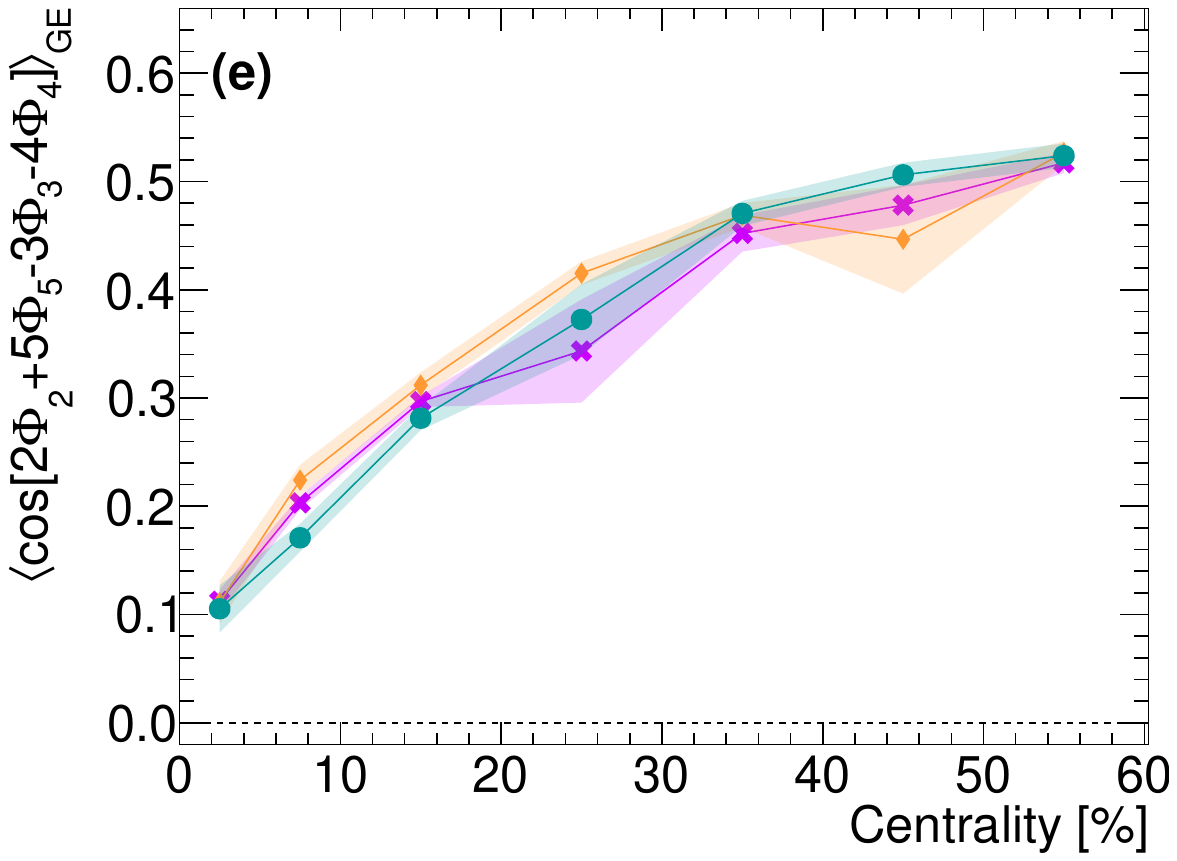}
    \includegraphics[width = 0.32\linewidth]{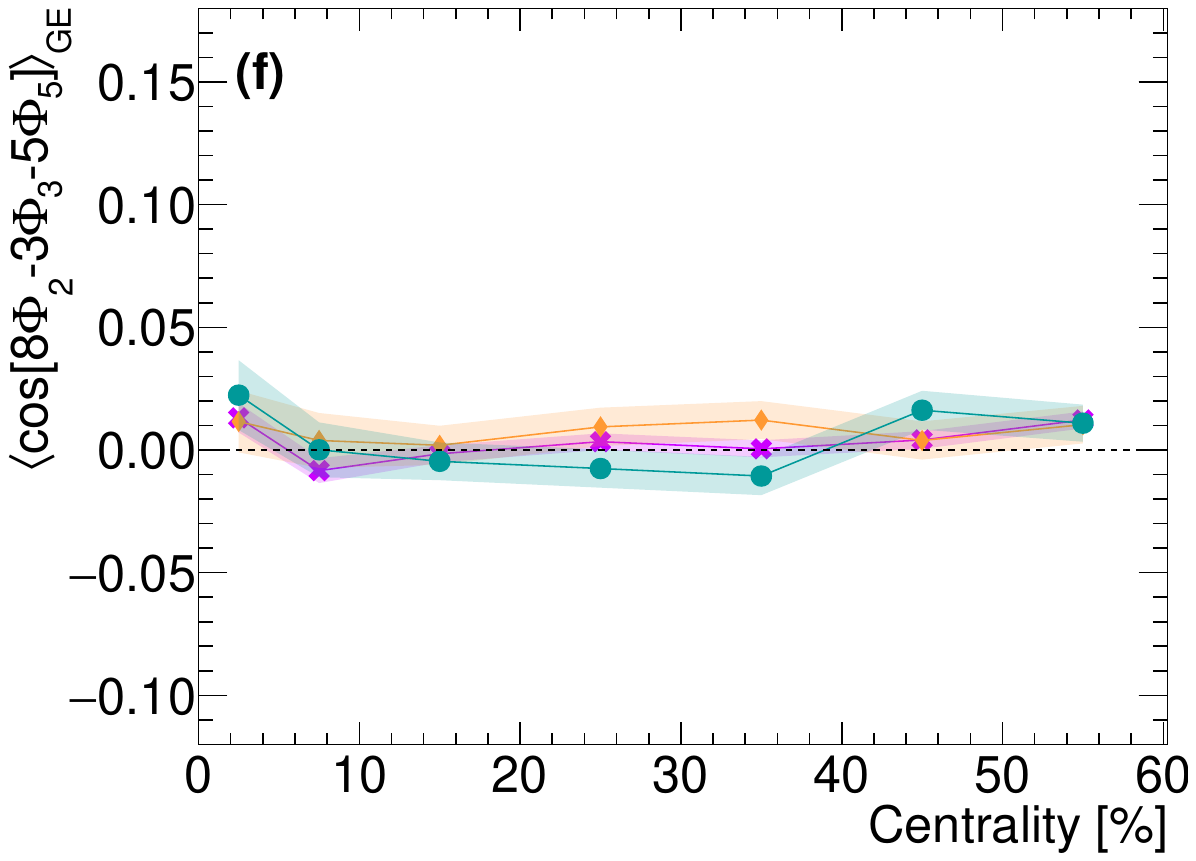}
    \caption{Centrality dependence of SPCs (shown in (a), (b), and (c)) among $n=2, 3, 5$ and corresponding PPCs (shown in (d), (e), and (f)) in Pb-Pb collisions at $\sqrt{s_{\rm NN}}=5.02$ TeV using AMPT. The calculation of SPCs considers the charged hadrons with $0.2<p_{\rm T}<5.0$ GeV/c, $|\eta|<0.8$.}
    \label{fig:SPCs235}
\end{figure*}

\subsubsection{SPCs of harmonic orders $n=2,3,5$}
\label{sec23and5}
Whereas a pentagonal symmetry favors an approximate triangular symmetry (not equilateral, but isosceles triangles that can possibly be constructed using the vertices of a regular pentagon), the presence of elliptical symmetry is expected to be largely independent of either of the other two symmetries, when higher order anisotropies (e.g. $v_{30}$) are not dominant. 
However, the SPCs are also expected to arise from the cumulant expansion of the harmonics, where the higher-order harmonics have nonlinear contributions from the lower-order harmonics~\cite{Teaney:2010vd, Teaney:2012ke, Teaney:2013dta, ALICE:2024fus}. This explains the positive correlation the symmetry planes $\psi_a$, $\psi_b$, and $\psi_c$ for $c>b>a$ and $na+mb=c$ with $n,~m\in \mathbb{N} $. This is explicitly observed in the strong positive correlation in $\langle\cos(2\psi_2+3\psi_3-5\psi_5)\rangle_{\rm GE}$ (Fig. \ref{fig:SPCs235} (a)) which is absent in $\langle \cos[2\psi_2+4\psi_4-6\psi_3]\rangle_{\rm GE}$ (Fig. \ref{fig:SPCs24} (b)). It is to be noted that a weaker correlation for $\langle\cos[8\psi_2-3\psi_3-5\psi_5]\rangle_{\rm GE}$ than $\langle\cos[2\psi_2+3\psi_3-5\psi_5]\rangle_{\rm GE}$ is attributed to the higher number of particle correlations in the former case.
\newline

For most of the correlators, it can be observed that the transformation from PPCs to SPCs is non-linear in nature. The non-linear response can be attributed to the following factors.
\begin{enumerate}
    \item As one moves from central to peripheral collisions, the number of participants (and parton density) decreases, which makes it difficult to transform initial PPCs to SPCs.
    \item Towards the peripheral collisions, due to a smaller number of participants and charged particle multiplicities, the event-by-event fluctuations dominate the values of PPCs. A non-linear contribution of these lower order PPCs to the higher order SPCs can induce a overall non-linear response of initial PPCs to final state SPCs.
    \item The Gaussian estimator, although better than other estimators of SPCs, is still not robust to a change in multiplicity.
\end{enumerate}

 A similar non-linear response of the system to the evolution of participant eccentricities to the final state anisotropic flow coefficients is also shown in Refs.~\cite{ATLAS:2015qwl, Prasad:2025ezg}.

\subsection{Spherocity dependence of SPCs and PPCs}
In Figs.~\ref{fig:SPCs236},~\ref{fig:SPCs24}, and~\ref{fig:SPCs235}, for most of the SPCs, regardless of their centrality, it is observed that the correlators for low-$S_{0}$ events have a higher magnitude of the correlation compared to $S_{0}$-int events. This is also true for the cases where the SPCs are driven by the elliptic geometry directly or indirectly arising in the cumulant expansion of the higher-order harmonics. This is because, in low-$S_0$ events, the azimuthal angles of a larger number of emergent particles are aligned to a particular direction in momentum space, which is characterised by the parametric direction $\hat{n}$ in Eq.~\ref{sphero}, than in the case of an event with high-$S_0$ of similar multiplicity. Therefore, in comparison to other events, since the azimuthal angles of the emergent particles are closer together in the momentum space for a low $S_0$ event, the combination $\{n_i\}_{i=1}^{l}$ where $\sum_{i=1}^ln_i = 0$, ensures that the argument that goes in as phase into the complex exponential in the RHS of Eq.~\eqref{corrln_def} is closer to zero. This implies that the low-$S_0$ events have dominating elliptic symmetry, and higher order harmonics, including triangular flow, contribute negligibly. The elliptic symmetry in the azimuthal distribution of particles ensures a vanishing sine term and a large cosine term in the RHS of Eq.~\eqref{corrln_def}. This leads to larger positive real parts. Hence, all the numerator terms during event averaging using only low-$S_{0}$ events in Eq.~\eqref{main_eqn} will have values that are more positive in comparison to those where the event averaging is done considering all shapes of events. Thus, if an SPC computed using events of all shapes is positive in a centrality class, it would mean that the same SPC would have a higher magnitude when computed using low-$S_{0}$ events.\newline

On the other hand, when SPCs for the high-$S_0$ (isotropic) events are considered, we have that the azimuthal angles are uniformly distributed, and hence the magnitude of SPCs evaluated using high-$S_0$ events would be smaller than that of the $S_0$-int case. Apart from this triviality in terms of magnitude, no particular behaviour in terms of whether an SPC evaluated from isotropic events in a particular centrality class - if it would be more positive or negative in comparison to the SPC obtained by considering events of all shapes - can be deduced solely from the mathematical construct of the correlators.\newline

A deviation from this trivial behaviour would be indicative of interesting underlying phenomena that lead to the observed geometrical distribution of particles in the transverse momentum space. The SPCs that deviate from their trivially expected values when computed using only low-$S_0$ or high-$S_0$ events are $\langle\cos[6(\psi_6-\psi_3)]\rangle_{\rm GE}$, $\langle\cos(2\psi_2+4\psi_4-6\psi_3)\rangle_{\rm GE}$, $\langle\cos[6(\psi_2-\psi_3)]\rangle_{\rm GE}$ and $\langle\cos(8\psi_2-3\psi_3-5\psi_5)\rangle_{\rm GE}$.\newline

The participant planes are determined by the initial geometry of the colliding system.  These are independent of the later stages of the evolution of the collision system. Hence, the participant planes are expected to be independent of the spherocity of an event, which largely takes its value due to the later stages of evolution. Consequently, any correlation between the participant planes is also expected to be independent of the spherocity classification of an event. This is exactly what is observed in all the PPCs, which are shown in Figs.~\ref{fig:SPCs24}(d)-(f), Figs.~\ref{fig:SPCs236}(d)-(f), Figs.~\ref{fig:SPCs235}(d)-(f). The values of the PPCs match within errors for the averages that have been computed using the different kinds of event shapes. Their dependence on centrality is similar to that of the corresponding SPCs in Figs.~\ref{fig:SPCs236}(a)-(c), Figs.~\ref{fig:SPCs24}(a)-(c), Figs.~\ref{fig:SPCs235}(a)-(c). \newline

For PPCs that grow with decreasing centrality, shown in Figs.~\ref{fig:SPCs236} (e), Fig.~\ref{fig:SPCs24} (d) and (f), and Figs.~\ref{fig:SPCs235} (d) and (e), a striking difference can be seen between their spherocity dependence.
A noticeable feature is that the SPCs mimic qualitatively the centrality dependence of the PPCs for all correlations, at least up to mid-central events (40-50\% most central). Also, there are some other SPCs, e.g, $\langle\cos[6(\psi_6-\psi_2)]\rangle_{\rm GE}$ and $\langle\cos(2\psi_2+4\psi_4-6\psi_6)\rangle_{\rm GE}$ for which the SPCs that are computed using low-$S_0$ events in the respective centrality class, mimic the qualitative centrality dependence of their corresponding PPCs, even for peripheral events. \newline

The general behaviour of the increase of these PPCs as well as SPCs, when we move towards mid-central event classes, is that higher-order anisotropies start playing a dominant role. Hence, the lower-order symmetry planes get correlated to each other non-trivially. For most central events, which are isotropic in transverse momentum space, there need not be any correlation between the different harmonic orders of the symmetry planes, i.e., the orientations of the symmetry planes need not be correlated among themselves at all. This is exactly what is observed in both the SPCs and the PPCs, as shown in Fig.~\ref{fig:SPCs236} (e), Fig.~\ref{fig:SPCs24} (d) and (f), and Figs.~\ref{fig:SPCs235} (d) and (e). \newline

Another comparison of SPCs to their respective PPCs can be made in Fig.~\ref{fig:SPCs24}(e) and Fig.~\ref{fig:SPCs236}(d), for PPCs that decrease as we move towards peripheral collisions. The reason for which $\langle\cos[6(\Phi_3-\Phi_6)]\rangle_{\rm GE}$ decreases with centrality is due to its geometrical detail, as has been pointed out in Section~\ref{236}. For the PPC between the harmonic orders $n=2,3$ and 4, we have the fact that in most central collisions, the contribution of $v_2$ to anisotropic flow is less in comparison to that in peripheral events. As $\psi_2$ and $\psi_4$ together have an appreciable correlation with $\psi_6$ (Figs.~\ref{fig:SPCs24}(d) and (f)) and $\psi_3$ too has an appreciable correlation with $\psi_6$ (Fig.~\ref{fig:SPCs236}(d)), we observe a similar trend with centrality, for the simplest possible combination that has been constructed for the PPCs corresponding to $n=2,3$ and 4. \newline

In addition to the SPCs and PPCs discussed above, Fig.~\ref{fig:SPCs236}(f) and Fig.~\ref{fig:SPCs235}(f) show the comparison of SPCs to the corresponding PPCs, which are closer to zero within error bars across all centrality classes. These correlations have very small magnitudes and vary randomly about zero. This trend applies to their respective SPCs as well. However, in these cases, it is observed that the SPCs have a higher magnitude in comparison to the PPCs, even though their variation is random about the $x-$axis. In addition, no particular behaviour in terms of event shape can be observed for the PPCs and SPCs shown in Fig.~\ref{fig:SPCs236}(c),(f) and Fig.~\ref{fig:SPCs235}(c),(f). This is an indication that the correlator is actually close to zero, irrespective of the event shape chosen, and hence the SPCs obtained using averaging over pencil-like or isotropic events show no specific centrality behaviour. This behaviour of the SPCs has been reported in the ALICE experiment \cite{ALICE:2023wdn}. In all the various comparisons that were made in this section, between PPCs and their corresponding SPCs, we observe a clear qualitative match of their centrality dependence. This shows, thus, that the SPCs are a good probe for PPCs.

\section{Summary and Conclusion}
\label{sec_summary}
In summary, we study the centrality and transverse spherocity dependence of symmetry plane correlations in Pb-Pb collisions at $\sqrt{s_{\rm NN}}=5.02$ TeV using AMPT. For the first time, this paper presents a study of transverse spherocity dependence of SPCs using Gaussian estimators within the framework of AMPT. The Gaussian estimator of SPCs has been observed to be strongly sensitive to event shape, and their variation with spherocity can be used to study them in greater detail in different centrality classes of interest, as per the physics process being addressed. When SPCs are computed using isotropic events, they have been observed to have smaller magnitude and random variation about zero, showing no particular centrality dependence, except for $\langle\cos[4(\psi_4-\psi_2)]\rangle_{\rm GE}$, where, it can be distinctly observed that the SPC obtained using isotropic events only takes on negative values, irrespective of the centrality. \newline

Most of the SPCs, except for those that are very close to zero, when computed for the low-$S_0$ events, are all larger than the $S_0$-int case. Thus, these will be particularly useful in examining if a certain SPC is actually zero due to the underlying geometry of the system or if it is zero simply due to it being computed from the chosen sample of events with all kinds of event shapes. Symmetry planes of even orders are likely to have greater correlation, as has been seen for the SPCs of $n=2,4,6$. All the SPCs, especially for the low-$S_0$ case, qualitatively mimic the centrality dependence of their corresponding PPCs. The PPCs do not depend strongly on spherocity, as they are not affected by the several processes that ultimately lead spherocity to take on the value that it does for an event. The qualitative estimation of PPCs by the SPCs will be an important tool to determine the relevant final state effect of nuclear structure at high-energy heavy-ion collisions, such as in $^{16}$O-$^{16}$O collisions at the RHIC and LHC \cite{Zhao:2017yhj}.

\section*{Acknowledgment}
S. T. acknowledges the KVPY SB scholarship. S. P. acknowledges the University Grants Commission (UGC), Government of India, for doctoral fellowship. R.~S. sincerely acknowledges the DAE-DST, Government of India, funding under the mega-science project – “Indian participation in the ALICE experiment at CERN” bearing Project No. SR/MF/PS-02/2021-IITI (E-37123). The authors gratefully acknowledge the computing facility provided by the Grid Computing Facility, VECC, Kolkata, India.

\bibliographystyle{ieeetr}

\end{document}